\documentclass[11pt]{article}
\usepackage{amsmath,amssymb}
\usepackage{epsfig,graphics,graphicx}
\usepackage{color,multicol}
\usepackage{natbib}
\usepackage{multirow}
\usepackage{color, colortbl}
\definecolor{Gray}{gray}{0.9}
\usepackage{setspace}
\begin{document}

\title{Modeling for Dynamic Ordinal Regression Relationships: An Application to Estimating Maturity of Rockfish in California}
\author{Maria DeYoreo and Athanasios Kottas
\thanks{
M. DeYoreo (maria.deyoreo@stat.duke.edu) is postdoctoral researcher, Department of Statistical Science, 
Duke University, Durham, NC, 27708, USA, and A. Kottas (thanos@soe.ucsc.edu) is Professor, Department 
of Applied Mathematics and Statistics, University of California, Santa Cruz, CA, 95064, USA.
The authors wish to thank Stephan Munch for providing the rockfish data and for several useful comments 
on the interpretation of the results, as well as Alec MacCall, Don Pearson, and Marc Mangel for valuable
input on data collection and on key aspects of the specific problem from fisheries research.
This research is part of the Ph.D dissertation of M. DeYoreo, completed at University of California, Santa Cruz, 
and was supported in part by the National Science Foundation under award DMS 1310438. M. DeYoreo is currently supported by a grant from the National Science Foundation under award SES 1131897.}
}

\date{}

\maketitle
\begin{abstract}

\noindent
We develop a Bayesian nonparametric framework for modeling ordinal regression relationships 
which evolve in discrete time. The motivating application involves a key problem in fisheries research 
on estimating dynamically evolving relationships between age, length and maturity, the latter recorded 
on an ordinal scale. The methodology builds from nonparametric mixture modeling for the joint stochastic 
mechanism of covariates and latent continuous responses. This approach yields highly flexible inference 
for ordinal regression functions while at the same time avoiding the computational challenges of parametric 
models. A novel dependent Dirichlet process prior for time-dependent mixing distributions extends the 
model to the dynamic setting. The methodology is used for a detailed study of relationships between 
maturity, age, and length for Chilipepper rockfish, using data collected over 15 years along the coast 
of California.

\end{abstract}


\newpage

\section{Introduction}
Consider ordinal responses collected along with covariates over discrete time. Furthermore, assume multiple observations are recorded at each point in time. This article develops Bayesian nonparametric modeling and inference for a discrete time series of ordinal regression relationships. Our aim is to provide flexible inference for the series of regression functions, estimating the unique relationships present at each time, while introducing dependence by assuming each distribution is correlated with its predecessors.

Environmental characteristics consisting of ordered categorical and continuous measurements may be monitored and recorded at different points in time, requiring a model for the temporal relationships between the environmental variables. The relationships present at a particular point in time are of interest, as well as any trends or changes which occur over time. Empirical distributions in environmental settings may exhibit non-standard features including heavy tails, skewness, and multimodality. To capture these features, one must move beyond standard parametric models in order to obtain more flexible inference and prediction. 

\begin{figure}[t!]
\centering
\includegraphics[height=3.7in,width=5.3in]{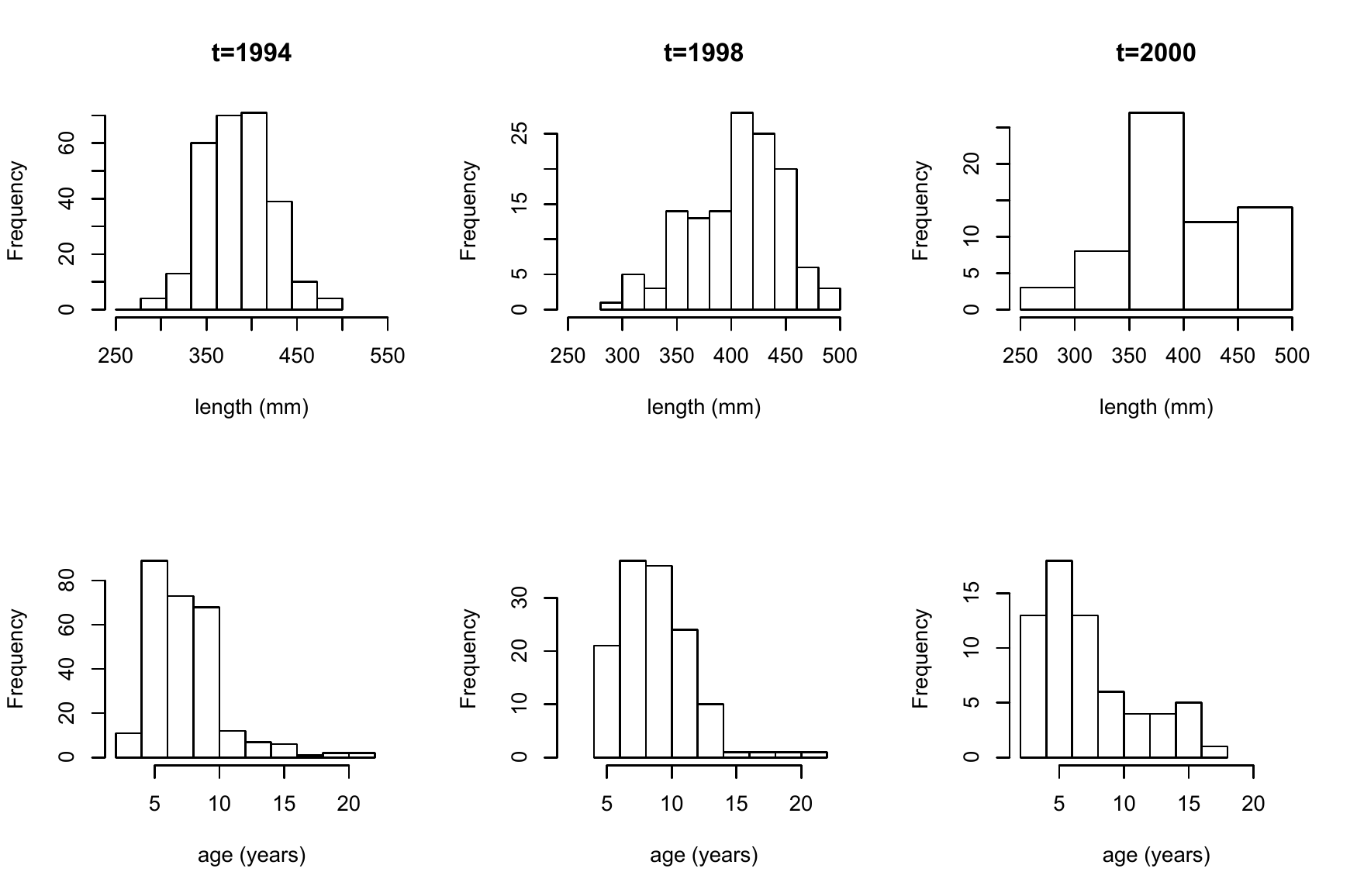}
\caption{Data histograms of length and age for female Chilipepper rockfish
in years $1994$, $1998$, and $2000$.}
\label{fig:length_hist}
\end{figure}

\begin{figure}[t!]
\centering
\includegraphics[height=5.5in,width=5in]{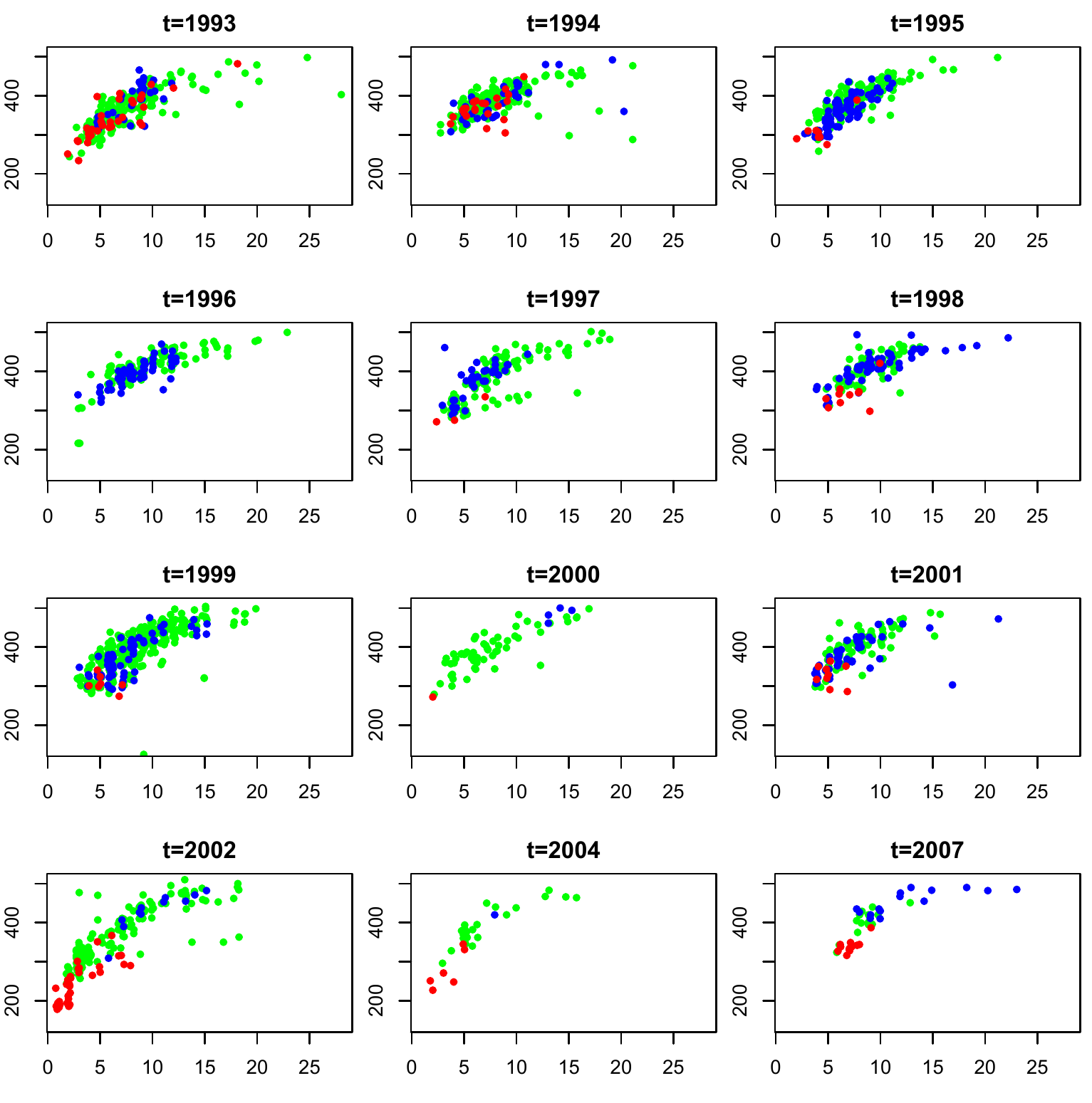}
\caption{Bivariate plots of length versus age at each year of data, with data points colored according 
to maturity level. Red represents level $1$, green level $2$, and blue level $3$. Values of age have been 
jittered to make the plots more readable.} 
\label{fig:biv_data}
\end{figure}

The motivating application for this work lies in modeling fish maturity as a function of age and length. This is a
key problem in fisheries science, one reason being that estimates of age at maturity play an important role in population estimates of sustainable harvest rates \citep{clark,hannahfish}. The specific data set comes from the National 
Marine Fisheries Service and consists of year of sampling, age recorded in years, length in millimeters, and maturity
for female Chilipepper rockfish, with measurements collected over 15 years along the coast of California. 
Maturity is recorded on an ordinal scale, with values taken to be from $1$ through $3$, where $1$ indicates 
immature and $2$ and $3$ represent pre- and post-spawning mature, respectively. More details on the data 
are provided in Section \ref{sec:application}. Exploratory analysis suggests both symmetric, unimodal 
as well as less standard shapes for the marginal distributions of length and age; histograms for 
three years are shown in Figure \ref{fig:length_hist}. Bivariate data plots of age and length suggest similar 
shapes across years, with some differences in location and scale, and clear differences in sample size, 
as can be seen from Figure \ref{fig:biv_data}. To make the plot more readable, random noise has been added 
to age, which is recorded on a discretized scale. Maturity level is also indicated; red color represents immature, 
green pre-spawning mature, and blue post-spawning mature. Again, there are similarities including the 
concentration of immature fish near the lower left quadrants, but also differences such as the lack of 
immature fish in years $1995$ through $2000$ as compared to the early and later years.

In addition to studying maturity as a function of age and length, inference for the age and length distributions 
is also important. This requires a joint model which treats age and length as random in addition to maturity. 
We are not aware of any existing modeling strategy for this problem which can handle multivariate mixed 
data collected over time. Compromising this important aspect of the problem, and assuming the regression 
of maturity on body characteristics is the sole inferential objective, 
a possible approach would be to use an ordered probit regression model. Empirical (data-based) estimates 
for the trend in maturity as a function of length or age indicate shapes which may not be captured well 
by a parametric model. For instance, the probability a fish is immature (level 1) is generally decreasing 
with length, however, in some of the years, the probability a fish is post-spawning mature (level 3) is 
increasing up to a certain length value and then decreasing. This is not a trend that can be captured by 
parametric models for ordinal regression \citep[][discuss some of these properties]{boes}. One could 
include higher order and/or interaction terms, though it is not obvious which terms to include, and 
how to capture the different trends across years.

In practice, virtually all methods for studying maturity as a function of age and/or length use logistic 
regression or some variant, often collapsing maturity into two levels (immature and mature) and treating 
each covariate separately in the analysis \citep[e.g.,][]{hannahfish,bobko}. \citet{bobko} applied 
logistic regression with length as a covariate, and to obtain an estimate of age at $50\%$ maturity 
(the age at which $50\%$ of fish are mature), they used their estimate for length at $50\%$ maturity 
and solved for the corresponding age given by the von Bertalanffy growth curve, which relates age to 
length using a particular parametric function. Others assume that maturity is independent of length 
after conditioning on age, leading to inaccurate estimates of the proportion mature at a particular 
age or length \citep{morgan}.

We develop a nonparametric Bayesian model to study time-evolving relationships between maturation, 
length, and age. These three variables constitute a random vector, and although maturity is recorded on 
an ordinal scale, it is natural to conceptualize an underlying continuous maturation variable. Distinguishing 
features of our approach include the joint modeling for the stochastic mechanism of maturation and 
length and age, and the ability to obtain flexible time-dependent inference for multiple ordinal maturation 
categories. While estimating maturity as a function of length and age is of primary interest, the joint 
modeling framework provides inference for a variety of functionals involving the three body characteristics.

Our goal is to construct a modeling framework for dynamic ordinal regression which avoids strict parametric 
assumptions and possesses features that make it well-suited to the fish maturity application, as well as to 
similar evolutionary biology problems on studying natural selection characteristics (such as survival or 
maturation) in terms of phenotypic traits.
To this end, we build on previous work on ordinal regression not involving time \citep{deyoreokottas}, 
where the ordinal responses arise from latent continuous variables, and the joint latent 
response-covariate distribution is modeled using a Dirichlet process (DP) mixture of multivariate normals 
\citep{muller}. In the context of the rockfish data, we model maturity, length, and age jointly, using a 
DP mixture. This modeling approach is further developed here to handle ordinal regressions which are 
indexed in discrete time, through use of a new dependent Dirichlet process (DDP) prior \citep{maceachern1999,maceachern2000}, which estimates the regression relationship at each time 
point in a flexible way, while incorporating dependence across time.

We review the model for ordinal regression without the time component in Section \ref{sec:ord_reg}. 
Section \ref{sec:ddp} introduces the DDP, and in Section \ref{sec:new_ddp}, we develop a new method 
for incorporating dependence in the DP weights to handle distributions indexed in discrete time. 
The model is then developed further in Section \ref{sec:application} in the context of the motivating 
application, and applied to analyze the rockfish data discussed above. Section \ref{sec:disc} concludes 
with a discussion. Technical details on properties of the DDP prior model, and on the posterior 
simulation method are included in the appendixes.

\section{Modeling Framework} \label{sec:methods}

\subsection{Bayesian Nonparametric Ordinal Regression} 
\label{sec:ord_reg}

We first describe our approach to regression in the context of a single distribution, that is, without any aspect of time. Let $\{(y_{i},\boldsymbol{x}_{i}): i=1,\dots,n\}$ denote the data, where each observation consists of an ordinal response $y_i$ along with a vector of covariates $\boldsymbol{x}_i=(x_{i1},\dots,x_{ip})$. The methodology is developed in \cite{deyoreokottas} for multivariate ordinal responses, however we work with a univariate response for notational simplicity and because this is the relevant setting for the application of interest.  Our model assumes the ordinal responses arise as discretized versions of latent continuous responses, which is natural for many settings and particularly relevant for the fish maturity application, as maturation is a continuous variable recorded on a discrete scale.  With $C$ categories, introduce latent continuous 
responses $(Z_1,\dots,Z_n)$ such that $Y_i=j$ if and only if $Z_i\in (\gamma_{j-1},\gamma_j]$, for $j=1,\dots,C$, 
and cut-offs $-\infty=\gamma_0<\gamma_1<\dots<\gamma_C=\infty$.

We focus on settings in which the covariates may be treated as random, which is appropriate, indeed necessary, for many environmental applications. In the fish maturity example, the body characteristics are interrelated and arise in the form of a data vector, and we are interested in various relationships, including but not limited to the way in which maturity varies with age and length. This motivates our focus on building a flexible model for the joint density $f(z,\boldsymbol{x})$, for which we apply a DP mixture of multivariate normals: $(z_i,\boldsymbol{x}_i)\mid G \stackrel{iid}{\sim} \int \mathrm{N}(\cdot;\boldsymbol{\mu},\boldsymbol{\Sigma})dG(\boldsymbol{\mu},\boldsymbol{\Sigma})$, with $G\mid\alpha,G_0\sim \mathrm{DP}(\alpha,G_0)$. 

By the constructive definition of the DP \citep{sethuraman}, a realization $G$ from a $\mathrm{DP}(\alpha,G_{0})$ is almost surely of the form $G=\sum_{l=1}^{\infty}p_l \delta_{\boldsymbol{\theta}_l}$.  The locations $\boldsymbol{\theta}_l=$ $(\boldsymbol{\mu}_l,\boldsymbol{\Sigma}_l)$ are independent realizations from the centering distribution $G_0$, and the weights are determined through stick-breaking from beta distributed random variables. In particular, let $v_l\stackrel{iid}{\sim}\text{beta}(1,\alpha)$, $l=1,2,\dots$, independently of $\{\boldsymbol{\theta}_l\}$, and define $p_1=v_1$, and for $l=2,3,\dots$, $p_l=v_l\prod_{r=1}^{l-1}(1-v_r)$. Therefore, the model for $f(z,\boldsymbol{x})$ has an almost sure representation as a countable mixture of multivariate normals, and implies the following induced model for the regression function:
 \begin{equation}\label{eqn:reg_fcn}\mathrm{Pr}(Y=j\mid  \boldsymbol{x};G)=\sum_{r=1}^{\infty}w_r(\boldsymbol{x})\int_{\gamma_{j-1}}^{\gamma_{j}}\text{N}(z;m_r(\boldsymbol{x}),s_r) \, dz
\end{equation}
 with covariate-dependent weights $w_r(\boldsymbol{x})\propto p_r \text{N}(\boldsymbol{x};\boldsymbol{\mu}_r^{x},\boldsymbol{\Sigma}_r^{xx})$, covariate-dependent means $m_r(\boldsymbol{x})=\mu_r^{z}+\boldsymbol{\Sigma}_r^{zx}(\boldsymbol{\Sigma}_r^{xx})^{-1}(\boldsymbol{x}-\boldsymbol{\mu}_r^{x})$, and variances $s_r=\Sigma_r^{zz}-\boldsymbol{\Sigma}_r^{zx}(\boldsymbol{\Sigma}_r^{xx})^{-1}\boldsymbol{\Sigma}_r^{xz}$. 
Here, $\boldsymbol{\mu}_r$ is partitioned into ${\mu}_r^{z}$ and $\boldsymbol{\mu}_r^{x}$ according to $Z$ 
and $\boldsymbol{X}$, and $({\Sigma}_r^{zz},\boldsymbol{\Sigma}_r^{xx},\boldsymbol{\Sigma}_r^{zx}, \boldsymbol{\Sigma}_r^{xz})$ are the components of the corresponding partition of covariance matrix $\boldsymbol{\Sigma}_r$.

This modeling strategy allows for non-linear, non-standard relationships to be captured, and overcomes many 
limitations of standard parametric models. In addition, the cut-offs may be fixed to arbitrary increasing values 
(which we recommend to be equally spaced and centered at zero) without sacrificing the ability of the model to 
approximate any distribution for mixed ordinal-continuous data. In particular, it can be shown that the induced 
prior model on the space of mixed ordinal-continuous distributions assigns positive probability to all 
Kullback-Leibler neighborhoods of any distribution in this space. This represents a key computational advantage 
over parametric models. We refer to \citet{deyoreokottas} for more details on model properties, a review of 
existing approaches to ordinal regression, and illustration of the benefits afforded by the nonparametric joint 
model over standard methods. This discussion refers to ordinal responses with three or more categories. For the 
case of binary regression, i.e., when $C=2$, additional restrictions are needed on the covariance matrix 
$\boldsymbol{\Sigma}$ to facilitate identifiability \citep{deyoreobinary}. 


\subsection{Dependent Dirichlet Processes}
\label{sec:ddp}

In developing a model for a collection of distributions indexed in discrete time, we seek to build on previous knowledge, retaining the powerful and well-studied DP mixture model marginally at each time $t\in\mathcal{T}$, with $\mathcal{T}=\{1,2,\dots\}$. We thus seek to extend the DP prior to model $G_\mathcal{T}=\{G_t:t\in\mathcal{T}\}$, a set of dependent distributions such that each $G_t$ follows a DP marginally. The dynamic DP extension can be developed by introducing temporal dependence in the weights  and/or atoms of the constructive definition, $G=$ 
$\sum_{l=1}^\infty p_l\delta_{\boldsymbol{\theta}_l}$.

The general formulation of the DDP introduced by \cite{maceachern1999,maceachern2000} expresses the atoms $\boldsymbol{\theta}_{l\mathcal{S}}=\{\boldsymbol{\theta}_{l,t}:t\in S\}$, $l=1,2,\dots$ as independently and identically distributed (i.i.d.) sample paths from a stochastic process over $S$, and the latent beta random variables which drive the weights, $\boldsymbol{v}_{l\mathcal{S}}=\{v_{l,t}:t\in S\}$, $l=1,2,\dots$, as i.i.d. realizations from a stochastic process with beta$(1,\alpha_t)$ marginal distributions. The distributions could be indexed in time, space, or by covariates, and $S$ represents the corresponding index set, being $\mathbb{Z}^+$ in our case. The DDP model for distributions indexed in discrete time expresses $G_{t}$ as $\sum_{l=1}^\infty p_{l,t}\delta_{{\boldsymbol{\theta}}_{l,t}}$, for $t\in \mathcal{T}$. The locations $\boldsymbol{\theta}_{l\mathcal{T}}=\{\boldsymbol{\theta}_{l,t}: t\in \mathcal{T}\}$ are i.i.d. for $l=1,2,\dots$, from a time series model for the kernel parameters. The stick-breaking weights $\boldsymbol{p}_{l\mathcal{T}}=\{p_{l,t}: t\in \mathcal{T}\}$, $l=1,2,\dots$, arise through a latent time series with beta$(1,\alpha_t)$ marginal distributions, independently of $\boldsymbol{\theta}_{l\mathcal{T}}$.

The general DDP can be simplified by introducing dependence only in the weights, such that the atoms are not time dependent, or alternatively, the atoms can be time dependent while the weights remain independent of time. We refer to these as common atoms and common weights models, respectively. While the majority of DDP applications fall into the common weights category, we believe the most natural of the two simplifications for time series is to assume that the locations are constant over time, and introduce dependence in the weights. Although we will end up working with the more general version of the DDP with dependent atoms and weights, for the application and related settings, it seems plausible that there is a fixed set of factors that determine the region in which the joint density of body characteristics is supported, but dynamics are caused by changes in the relative importance of the factors. The construction of dependent weights requires dependent beta random variables, so that $p_{1,t}=v_{1,t}$, $p_{l,t}=v_{l,t}\prod_{r=1}^{l-1}(1-v_{r,t})$, for $l=2, 3,\dots$, with each $\{v_{l,t}: t\in \mathcal{T}\}$ a realization from a time series model with beta($1,\alpha$) marginals. Equivalently, we can write $p_{1,t}=1-\beta_{1,t}$, $p_{l,t}=(1-\beta_{l,t})\prod_{r=1}^{l-1}\beta_{r,t}$, for $l=2, 3,\dots$, with each $\{\beta_{l,t}: t\in \mathcal{T}\}$ a realization from a time series model with beta($\alpha,1$) marginals. 

There have been many variations of the DDP model proposed in the literature. The common weights version was originally discussed by \citet{maceachern2000}, in which a Gaussian process was used to generate dependent locations, with the autocorrelation function controlling the degree to which distributions which are ``close'' are similar, and how quickly this similarity decays. \cite{diiorio} consider also a common weights model, in which the index of dependence is a covariate, a key application of DDP models. In the order-based DDP of \cite{griffin},  covariates are used to sort the weights. Covariate dependence is incorporated in the weights in the kernel and probit stick-breaking models of \cite{dunsonpark} and \cite{rodriguezdunson}, respectively, however these prior models do not retain the DP marginally. \cite{gelfandkottas} developed a DP mixture model 
for spatial data, using a spatial Gaussian process to induce dependence in DDP locations indexed by space. For data 
indexed in discrete time, \cite{rodriguezter} apply a common weights model, with atoms arising from a dynamic linear 
model. \citet{dilucca} develop a model for a single time series of continuous or binary responses through a DDP in 
which the atoms are dependent on lagged terms. \cite{XKS2015} construct a dynamic model for Poisson process 
intensities built from a DDP mixture with common weights and different types of autoregressive processes 
for the atoms. \cite{taddymarked} assumes the alternative simplification of the DDP with common atoms, and 
models the stick-breaking proportions $\{v_{l,t}: t\in \mathcal{T}\}$ using the positive correlated autoregressive 
beta process from \cite{mckenzie}. \cite{nieto} also use the common atoms simplification of the DDP, modeling 
a time series of random distributions by linking the beta random variables through latent binomially distributed 
random variables.

\subsection{A Time-Dependent Nonparametric Prior}
\label{sec:new_ddp}

To generate a correlated series $\{\beta_{l,t}: t\in \mathcal{T}\}$ such that each $\beta_{l,t}\sim \mathrm{beta}(\alpha,1)$ marginally, we define a stochastic process
\begin{equation}\label{eqn:beta_sp}
\mathcal{B}=\left\{\beta_t=\exp\left(-\frac{\zeta^2+\eta^2_t}{2\alpha}\right):\, t\in \mathcal{T}\right\},
\end{equation}
which is built from a standard normal random variable $\zeta$ and an independent stochastic process ${\eta}_{\mathcal{T}}=\{\eta_{t}: t\in \mathcal{T}\}$ with standard normal marginal distributions. This transformation leads to marginal distributions $\beta_{t}\sim \text{beta}(\alpha,1)$ for any $t$. To see this, take two independent standard normal random variables $Y_1$ and $Y_2$, such that $W=$ $(Y_1^2+Y_2^2)/2$ follows an exponential distribution
with mean 1, and thus $B = \exp(-W/\alpha)$ $\sim \mathrm{beta}(\alpha,1)$. 
To our knowledge, this is a novel construction for a common atoms DDP prior model. The practical utility 
of the transformation in (\ref{eqn:beta_sp}) is that it facilitates building the temporal dependence through 
Gaussian time-series models, while maintaining the DP structure marginally.

Because we work with distributions indexed in discrete time, we assume ${\eta}_{\mathcal{T}}$ to be a first-order autoregressive (AR) process, however alternatives such as higher order processes or Gaussian processes for spatially indexed data are  possible. The requirement of standard normal marginal distributions on ${\eta}_{\mathcal{T}}$ leads to a restriction on the variance of the AR(1) model, such that $\eta_{l,t}\sim \text{N}(\phi\eta_{l,t-1},1-\phi^2)$, $t=2,\dots,T$. Thus $|\phi|<1$, which implies stationarity for the stochastic process ${\eta}_{\mathcal{T}}$. Since $\mathcal{B}$ is a transformation of a strongly stationary stochastic process, it is also strongly stationary. Note that the correlation in $(\beta_{l,t},\beta_{l,t+k})$ is driven by the autocorrelation present in ${\eta}_{\mathcal{T}}$, and this induces dependence in the weights $(p_{l,t},p_{l,t+k})$, which leads to dependent distributions $(G_t,G_{t+k})$. 

We now explore this dependence, discussing some of the correlations implied by this prior model. First, consider the correlation of the beta random variables used to define the dynamic stick-breaking weights. Let $\rho_k=\text{corr}(\eta_t,\eta_{t+k})$, which is equal to $\phi^k$ under the assumption of an AR(1) process for $\eta_{\mathcal{T}}$. The autocorrelation function associated with $\mathcal{B}$ is 
\begin{equation}\label{eqn:acf_beta}
\text{corr}(\beta_t,\beta_{t+k}\mid \alpha,\phi) =
\frac{\alpha^{1/2}(1-\rho^2_k)^{1/2}(\alpha+1)^2(\alpha+2)^{1/2}}
{\left\{ (1-\rho^2_k+\alpha)^2-\alpha^2\rho^2_k \right\}^{1/2}}-\alpha(\alpha+2)
\end{equation}
as described in Appendix \ref{sec:app_corr}.  Smaller values for $\alpha$ lead to smaller correlations for any fixed $\phi$ at a particular lag, and $\phi$ controls the strength of correlation, with large $\phi$ producing large correlations which decay slowly. The parameters $\phi$ and $\alpha$ in combination can lead to a wide range of correlations, however $\alpha\geq1$ implies a lower bound near $0.5$ on the correlation for any lag $k$. In the limit, as $\alpha\rightarrow 0^+$, $\text{corr}(\beta_t,\beta_{t+k}\mid \alpha, \phi)\rightarrow 0$, and as $\alpha\rightarrow \infty$, $\text{corr}(\beta_t,\beta_{t+k}\mid \alpha, \phi)$ tends towards $0.5$ as $\rho_k\rightarrow 0^+$, and $1$ as $\rho_k\rightarrow 1^-$. Assuming $\rho_{k} = \phi^k$, gives $\lim_{\phi\to 1^-}\text{corr}(\beta_t,\beta_{t+k}\mid \alpha, \phi)=1$ and $\lim_{\phi \to 0^+}\text{corr}(\beta_t,\beta_{t+k}\mid \alpha, \phi)=\alpha^{1/2}(\alpha+1)(\alpha+2)^{1/2}-\alpha(\alpha+2).$ This tends upwards to $0.5$ quickly as $\alpha\rightarrow\infty$.

Note that $\text{corr}(\beta_t,\beta_{t+k}\mid \alpha, \phi)$ is a function of $\rho^2_k$ but not $\rho_k$, which is to be expected since $\eta_t$ enters the expression for $\beta_t$ only through $\eta_t^2$, and thus $-\rho_k$ and $\rho_k$ have the same effect in the correlation.  The same is true of the correlation in the DP weights, which is given below. 
We believe that $\phi\in(0,1)$ is a natural restriction, since we are building a stochastic process for distributions 
correlated in time through a transformation of an AR process, which intuition suggests should be positively 
correlated. However, all that is strictly required to preserve the DP marginals is $|\phi|<1$.

Assuming $\boldsymbol{\beta}_{l,\mathcal{T}}=\{\beta_{l,t}:t\in \mathcal{T}\}$ is generated by $\mathcal{B}$, 
from an underlying AR(1) process for ${\eta}_{\mathcal{T}}$ with coefficient $\phi$, we study the dependence 
induced in the resulting DDP weights at consecutive time points, $(p_{l,t},p_{l,t+1})$. 
The covariance is given by
{\begin{multline}\label{eqn:covw}
\text{cov}(p_{l,t},p_{l,t+1} \mid \alpha,\phi) =
\left\{\frac{\alpha^{3/2}(1-\phi^2)^{1/2}}{(2+\alpha)^{1/2}\left\{ (1-\phi^2+\alpha)^2-\alpha^2\phi^2 \right\}^{1/2}} \right\}^{l-1}\\ \left\{1-\frac{2\alpha}{\alpha+1} + \frac{\alpha^{3/2}(1-\phi^2)^{1/2}}{(2+\alpha)^{1/2}\left\{ (1-\phi^2+\alpha)^2-\alpha^2\phi^2 \right\}^{1/2}}\right\}
-\frac{\alpha^{2l-2}}{(1+\alpha)^{2l}},
\end{multline}}
which can be divided by 
$\mathrm{var}(p_{l,t}\mid \alpha)=$ $\{ 2 \alpha^{l-1}/((1+\alpha)(2+\alpha)^{l}) \}
- \{ \alpha^{2l-2}/(1+\alpha)^{2l} \}$
to yield $\text{corr}(p_{l,t},p_{l,t+1}\mid \alpha, \phi)$; note that $\text{E}(p_{l,t} \mid \alpha)=$
$\text{E}(p_{l,t+1} \mid \alpha)$ and $\text{var}(p_{l,t} \mid \alpha)=$ $\text{var}(p_{l,t+1} \mid \alpha)$.
Derivations are given in Appendix \ref{sec:app_corr}. 
This expression is decreasing in index $l$, and larger values of $\phi$ lead to larger correlations 
in the weights at any particular $l$. Moreover, the decay in correlations with weight index is faster for small 
$\alpha$ and small $\phi$. As $\alpha\rightarrow 0^+$, $\text{corr}(p_{1,t},p_{1,t+1}\mid \alpha, \phi)\rightarrow 1$ 
for any value of $\phi$, and as $\alpha\rightarrow \infty$, $\text{corr}(p_{1,t},p_{1,t+1}\mid \alpha, \phi)$ 
is contained in $(0.5,1)$, with values closer to $1$ for larger $\phi$. Note that 
$\text{corr}(p_{l,t},p_{l,t+k}\mid \alpha, \phi)$ has the same expression as 
$\text{corr}(p_{l,t},p_{l,t+1}\mid \alpha, \phi)$, but with $\phi$ replaced by $\phi^k$; it is 
thus decreasing with the lag $k$, with the speed of decay controlled by $\phi$.

Finally, assume $G_t$ is a random distribution on $\mathbb{R}^M$, such that $G_t=$
$\sum_{l=1}^\infty p_{l,t}\delta_{\boldsymbol{\theta}_l}$, where the $\{ p_{l,t} : t \in \mathcal{T} \}$ 
are defined through the dynamic stick-breaking weights $\{ \beta_{l,t} : t \in \mathcal{T} \}$, 
and the $\boldsymbol{\theta}_l$ are i.i.d. from a distribution $G_{0}$ on $\mathbb{R}^M$. 
Consider two consecutive distributions $(G_t,G_{t+1})$, and a measurable subset 
$A \subset \mathbb{R}^{M}$. The correlation of consecutive distributions, 
$\text{corr}(G_t(A),G_{t+1}(A) \mid \phi,\alpha,G_0)$, is discussed in Appendix \ref{sec:app_corr}.

\section{Estimating Maturity of Rockfish}
\label{sec:application}

\subsection{Chilipepper Rockfish Data}

We now utilize the method for incorporating dependence into the weights of the DP and the approach to ordinal regression involving DP mixtures of normals for the latent response-covariate distribution to further develop the DDP mixture modeling framework for dynamic ordinal regression. 

In the original rockfish data source, maturity is recorded on an ordinal scale from $1$ to $6$, representing immature (1), early and late vitellogenesis $(2,3)$, eyed larvae $(4)$, and post-spawning $(5,6)$. Because scientists are not necessarily interested in differentiating between every one of these maturity levels, and to make the model output simple and more interpretable, we collapse maturity into three ordinal levels, representing immature $(1)$, pre-spawning mature $(2,3,4)$, and post-spawning mature $(5,6)$. 

Many observations have age missing or maturity recorded as unknown. Exploratory analysis suggests there to be no systematic pattern in missingness. Further discussion with fisheries research scientists having expertise in aging of rockfish and data collection revealed that the reason for missing age in a sample is that otoliths (ear stones used in aging) were not collected or have not yet been aged. Maturity may be recorded as unknown because it can be difficult to distinguish between stages, and samplers are told to record unknown unless they are reasonably sure of the stage. Therefore, there is no systematic reason that age or maturity is not present, and it is reasonable to assume that the data are missing at random, or that the probability an observation is missing does not depend on the missing values, allowing us to ignore the missing data mechanism, and base inferences only on the complete data \citep[e.g.,][]{rubin,gcsr}.

Age can not be treated as a continuous covariate, as there are approximately $25$ distinct values of age in over $2,200$ observations. Age is in fact an ordinal random variable, such that a recorded age $j$ implies the fish was between $j$ and $j+1$ years of age. This relationship between discrete recorded age and continuous age is obtained by the following reasoning. Chilipepper rockfish are winter spawning, and the young are assumed to be born in early January. The annuli (rings) of the otiliths are counted in order to determine age, and these also form sometime around January. Thus, for each ring, there has been one year of growth. 

We therefore treat age much in the same way as maturity, using a latent continuous age variable. Let $U$ represent observed ordinal age, let $U^*$ represent underlying continuous age, and assume, for $j=1,2,\dots,$ that $U=j$ iff $U^*\in(j,j+1]$. Equivalently, $U=j$ iff $\log(U^*)\in (\log(j),\log(j+1)]$, for $j=1,2,\dots$, and $U=0$ iff $\log(U^*)\in (-\infty,0]$, so that the support of the latent continuous random variable corresponding to age is $\mathbb{R}$. Letting $W$ be the latent continuous random variable which determines $U$ through discretization, we assume $u_{t,i}=j$ iff $w_{t,i}\in(\log(j),\log(j+1)]$, for $j=0,1,\dots$, so that $W$ is interpretable as log-age on a continuous scale.

Considering year of sampling as the index of dependence, observations occur in years $1993$ through $2007$, indexed by $t=1,\dots,T=15$, with no observations in $2003$, $2005$, or $2006$. Let the missing years be given by $\boldsymbol{s}$, here $\boldsymbol{s}=\{11,13,14\}$, and let $\boldsymbol{s}^c=\{1,\dots,T\} \setminus \boldsymbol{s}$ represent all other years in $\{1,\dots,T\}$. This situation involving time points in which data is completely missing is not uncommon in these types of problems, and can be handled with our model for equally spaced time points. We retain the ability to provide inference and estimation in years for which no data was recorded, which we will see are reasonable and exhibit more uncertainty than in other years.

\subsection{Hierarchical Model and Implementation Details}
\label{sec:fish_methods}

The common atoms DDP model presented in Section \ref{sec:new_ddp} was tested extensively on simulated data, 
and performed very well in capturing trends in the underlying distributions when there were no missing years 
or forecasting was not the focus. In these settings, the common atoms model is sufficiently powerful from an inferential perspective such that the need to turn to a more complex model is diminished. However, as a consequence of the need to force the same set of components to be present at each time point, density estimates at missing time points tend to resemble an average across all time points, which is not desirable when a trend or change in support is suggested. A more general model is required for this application, since it is important to make inferences in years for which data was not collected. We therefore consider a simple extension, adding dependence through a vector autoregressive model in the mean component of the DP atoms, such that $\boldsymbol{\theta}_l=(\boldsymbol{\mu}_l,\boldsymbol{\Sigma}_l)$ becomes $\boldsymbol{\theta}_{l,t}=(\boldsymbol{\mu}_{l,t},\boldsymbol{\Sigma}_l)$.

Assuming $Z$ represents maturity $Y$ on a continuous scale, $W$ is interpretable as log-age, and $X$ represents length, a dependent nonparametric mixture model is applied to estimate the time-dependent distributions of the trivariate continuous random vectors $\boldsymbol{Y}^*_{ti}=(Z_{ti},W_{ti},X_{ti})$, for $t=1,\dots,T$, and $i=1,\dots,n_t$. In our notation, $T$ is the number of years or the final year containing data (and it is possible that not all years in $\{1,\dots,T\}$ contain observations), and $n_t$ is the sample size in year $t$.

We utilize the computationally efficient approach to inference which involves truncating the countable representation for each $G_t$ to a finite level $N$ \citep{ishjames}, such that the dependent stick-breaking weights are given by $p_{1,t}=1-\beta_{1,t}$, $p_{l,t}=(1-\beta_{l,t})\prod_{r=1}^{l-1}\beta_{r,t}$, for $r=1,\dots,N-1$, and $p_{N,t}=\prod_{l=1}^{N-1}\beta_{l,t}$, ensuring $\sum_{l=1}^N p_{l,t}=1$. Since $\alpha$ is not a function of $t$,
the same truncation level is applied for all mixing distributions. In choosing the truncation level, we use the expression relating $N$ to the expectation of the sum of the first $N$ weights $w_1,\dots,w_N$ generated from stick-breaking of beta$(1,\alpha)$ random variables. The expression is $\mathrm{E}(\sum_{j=1}^N w_j\mid \alpha)=$
$1-(\alpha/(\alpha+1))^N$, which can be further averaged over the prior for $\alpha$ to obtain 
$\mathrm{E}(\sum_{j=1}^N w_j)$, with $N$ chosen such that this expectation is close to 1 up to 
the desired level of tolerance for the approximation.
The hierarchical model can be expressed as follows:
\begin{eqnarray}\label{eqn:hier}
 y_{t,i}=j \leftrightarrow \gamma_{j-1}<z_{t,i}\leq \gamma_{j}, \, t\in \boldsymbol{s}^c, \, i=1,\dots,n_t \nonumber\\
 u_{t,i}=j \leftrightarrow \log(j)<w_{t,i}\leq \log(j+1), \, t\in \boldsymbol{s}^c, \, i=1,\dots,n_t \nonumber\\
\{\boldsymbol{y}^*_{t,i}\}\mid \{\boldsymbol{\mu}_{l,t}\},\{\boldsymbol{\Sigma}_l\},\{L_{t,i}\}\sim \prod_{t\in \boldsymbol{s}^c}\prod_{i=1}^{n_t}\text{N}(\boldsymbol{\mu}_{L_{t,i},t},\boldsymbol{\Sigma}_{L_{t,i}}) \nonumber \\
\{L_{t,i}\}\mid \{\eta_{l,t}\},\{\zeta_l\}\sim \prod_{t\in \boldsymbol{s}^c}\prod_{i=1}^{n_t}\sum_{l=1}^{N}p_{l,t}\delta_l(L_{t,i}) \nonumber \\
\zeta_l\stackrel{iid}{\sim} \text{N}(0,1),\quad l=1,\dots,N-1\nonumber \\
\eta_{l,1}\stackrel{iid}{\sim} \text{N}(0,1),\quad l=1,\dots, N-1\nonumber \\
\eta_{l,t}\mid \eta_{l,t-1},\phi \sim \text{N}(\phi\eta_{l,t-1},1-\phi^2),\quad l=1,\dots,N-1,\, t=2,\dots, T\nonumber \\
 \boldsymbol{\mu}_{l,1}\mid \boldsymbol{m_0},\boldsymbol{V_0}\sim \text{N}(\boldsymbol{m_0},\boldsymbol{V_0}),\quad l=1,\dots,N \nonumber \\
\boldsymbol{\mu}_{l,t}\mid \boldsymbol{\mu}_{l,t-1},\Theta,\boldsymbol{m},\boldsymbol{V}\sim \text{N}(\boldsymbol{m}+\Theta\boldsymbol{\mu}_{l,t-1},\boldsymbol{V}),\quad l=1,\dots,N,\, t=2,\dots, T \nonumber \\
\boldsymbol{\Sigma}_l\mid \nu,\boldsymbol{D} \stackrel{iid}{\sim}\text{IW}(\boldsymbol{\Sigma}_l;\nu,\boldsymbol{D}),\quad l=1,\dots,N
\end{eqnarray}
with priors on $\alpha$, $\boldsymbol{\psi}=(\boldsymbol{m},\boldsymbol{V},\boldsymbol{D})$, $\phi$, and $\Theta$. Recall that $\beta_{l,t}$ is defined through $(\eta_{l,t},\zeta_l,\alpha)$, and the $\{\beta_{l,t}\}$ determine the $\{p_{l,t}\}$ through stick-breaking.

The parameters $\{\zeta_l\}$ and $\{\eta_{l,t}\}$ can be updated individually with slice samplers, which involves alternating simulation from uniform random variables and truncated normal random variables, implying draws for $\{\beta_{l,t}\}$, and hence $\{p_{l,t}\}$. The configuration variables $L_{t,i}$ are drawn from discrete distributions on $\{1,\dots,N\}$, with probabilities proportional to $p_{l,t}\text{N}(\boldsymbol{y}^{*}_{t,i};\boldsymbol{\mu}_{l,t},\boldsymbol{\Sigma}_l)$ for $l=1,\dots,N$. The update for $\Sigma_l$ is $\text{IW}(\nu+M_l,\boldsymbol{D}+\sum_{\{(t,i):L_{ti}=l\}}(\boldsymbol{y}^{*}_{t,i}-
\boldsymbol{\mu}_{l,t})(\boldsymbol{y}^{*}_{t,i}-\boldsymbol{\mu}_{l,t})^T)$, where $M_l=|\{t,i\}:L_{t,i}=l|$, and each $\boldsymbol{\mu}_{l,t}$ is updated from a normal distribution. The parameters $\alpha$ and $\phi$, given priors $\text{IG}(a_\alpha,b_\alpha)$, and uniform on $(0,1)$ or $(-1,1)$, respectively, can be sampled using Metropolis-Hastings steps.  We assume that $\Theta$ is diagonal, with elements $(\theta_1,\theta_2,\theta_3)$, however we advocate for a full covariance matrix $\boldsymbol{V}$. This implies that each element of $\boldsymbol{\mu}_{l,t}$ has a mean which depends only on the corresponding element of $\boldsymbol{\mu}_{l,t-1}$, however there exists dependence in the elements of $\boldsymbol{\mu}_{l,t}$, which seems reasonable for most applications. We assume uniform priors on $(0,1)$ or $(-1,1)$ for each element of $\Theta$, and update them with a Metropolis-Hastings step. Finally, the parameters $\boldsymbol{\psi}$ have closed-form full conditional distributions, given priors $\boldsymbol{m}\sim \text{N}(\boldsymbol{a_m},\boldsymbol{B_m})$, $\boldsymbol{V}\sim \text{IW}(a_{\boldsymbol{V}},\boldsymbol{B_V})$, $\boldsymbol{D}\sim \text{W}(a_{\boldsymbol{D}},\boldsymbol{B_D})$.
The full conditionals and posterior simulation details are further described in Appendix \ref{sec:app_mcmc}.

To implement the model, we must specify the parameters of the hyperpriors on $\boldsymbol{\psi}$. A default specification strategy is developed by considering the limiting case of the model as $\alpha\rightarrow0^+$ and
$\Theta \rightarrow \boldsymbol{0}$, which results in a single normal distribution for $\boldsymbol{Y}^*_t$. 
In the limit, with $\boldsymbol{Y}^*_t\mid\boldsymbol{\mu}_t,\boldsymbol{\Sigma} \sim \text{N}(\boldsymbol{\mu}_t,\boldsymbol{\Sigma})$ and $\boldsymbol{\mu}_t\mid \boldsymbol{m},\boldsymbol{V} \sim \text{N}(\boldsymbol{m},\boldsymbol{V})$, we find $\text{E}(\boldsymbol{Y}^*_t)=\boldsymbol{a_m}$ and $\text{Cov}(\boldsymbol{Y}^*_t)=\boldsymbol{B_m}+\boldsymbol{B_V}(a_{\boldsymbol{V}}-d-1)^{-1}+a_{\boldsymbol{D}}\boldsymbol{B_D}(\nu-d-1)^{-1}$, where $d$ is the response-covariate dimension, here $d=3$. The only covariate information we require is an approximate center (such as the midpoint of the data) and range, denoted by $c^x$ and $r^x$ for $X$, and analogously for $U$. We use $c^x$ and $r^x/4$ as proxies for the marginal mean and standard deviation of $X$. We also seek to scale the latent variables appropriately. The centers  and ranges $c^u$ and $r^u$ provide approximate centers $c^w$ and ranges $r^w$ of latent log-age $W$. Since $Y$ is supported on $\{1,\dots,C\}$, latent continuous $Z$ must be supported on values slightly below $\gamma_1$ up to slightly above $\gamma_{C-1}$, so that $r^z/4$ is a proxy for the standard deviation of $Z$, where $r^z=(\gamma_{C-1}-\gamma_1)$. Using these mean and variance proxies, we fix $a_{\boldsymbol{m}}=(0,c^u,c^x)$. Each of the three terms in $\text{Cov}(\boldsymbol{Y}^*_t)$ can be assigned an equal part of the total covariance, for instance being set to $3^{-1}\text{diag}\{(r^z/4)^2,(r^w/4)^2,(r^x/4)^2\}$. For dispersed but proper priors, $\nu$, $a_{\boldsymbol{V}}$ and $a_{\boldsymbol{D}}$ can be fixed to small values such as $d+2$, and $\boldsymbol{B_m}$, $\boldsymbol{B_V}$, and $\boldsymbol{B_D}$ determined accordingly.

It remains to specify $\boldsymbol{m}_0$ and $\boldsymbol{V}_0$, the mean and covariance for the initial distributions $\boldsymbol{\mu}_{l,1}$. We propose a fairly conservative specification, noting that in the limit, $\text{E}(\boldsymbol{Y}^*_1)=\boldsymbol{m}_0$, and $\text{Cov}(\boldsymbol{Y}^*_1)=a_{\boldsymbol{D}}\boldsymbol{B_D}(\nu-d-1)^{-1}+\boldsymbol{V}_0$. Therefore, $\boldsymbol{m}_0$ can be specified in the same way as $\boldsymbol{a_m}$ but using only the subset of data at $t=1$, and $\boldsymbol{V}_0$ can be set to $\text{diag}\{(r^z_1/4)^2,(r^w_1/4)^2,(r_1^x/4)^2\}-a_{\boldsymbol{D}}\boldsymbol{B_D}(\nu-d-1)^{-1}$, where the subscript $1$ indicates the subset of data at $t=1$.

In simulation studies and the rockfish application, we observed a moderate to large amount 
of learning for all hyperparameters. For instance, for the rockfish data, the posterior distribution 
for $\phi$ was concentrated on values close to 1, indicating the DDP weights are strongly 
correlated across time. There was also moderate learning for $\alpha$ as its posterior distribution 
was concentrated around $0.5$, with small variance, shifted down relative to the prior 
which had expectation $2$. The posterior distribution for each element of $\boldsymbol{m}$ was
reduced in variance and concentrated on values not far from those indicated by the prior mean. 
The posterior samples for the covariance matrices $\boldsymbol{V}$ and $\boldsymbol{D}$
supported smaller variance components than suggested by the prior.

\subsection{Results}

Various simulation settings were developed to study both the common atoms version of the model 
and the more general version \citep[][chapter 4]{deyoreothesis}. 
%
%
While we focus only on the fish maturity data application in this paper, our extensive simulation
studies have revealed the inferential power of the model under different scenarios for the true
latent response distribution and ordinal regression relationships.

\begin{figure}
\centering
\includegraphics[height=3.5in,width=5.5in]{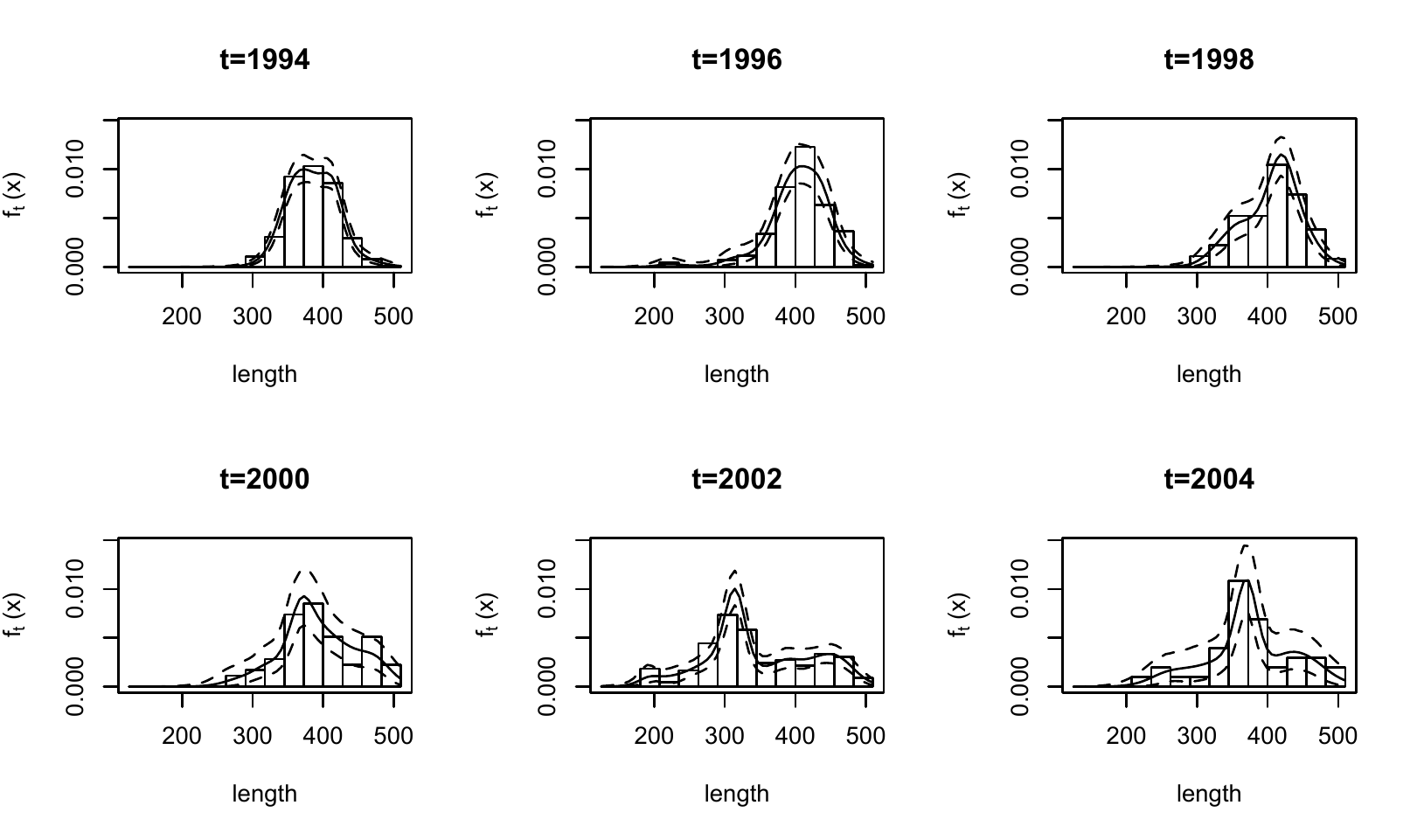}
\caption{Posterior mean and $95\%$ interval estimates for the density of length (in millimeters) 
across six years, with the data shown as a histogram.}
\label{fig:dens_length}
\end{figure}

\begin{figure}
\centering
\includegraphics[height=3.5in,width=5.5in]{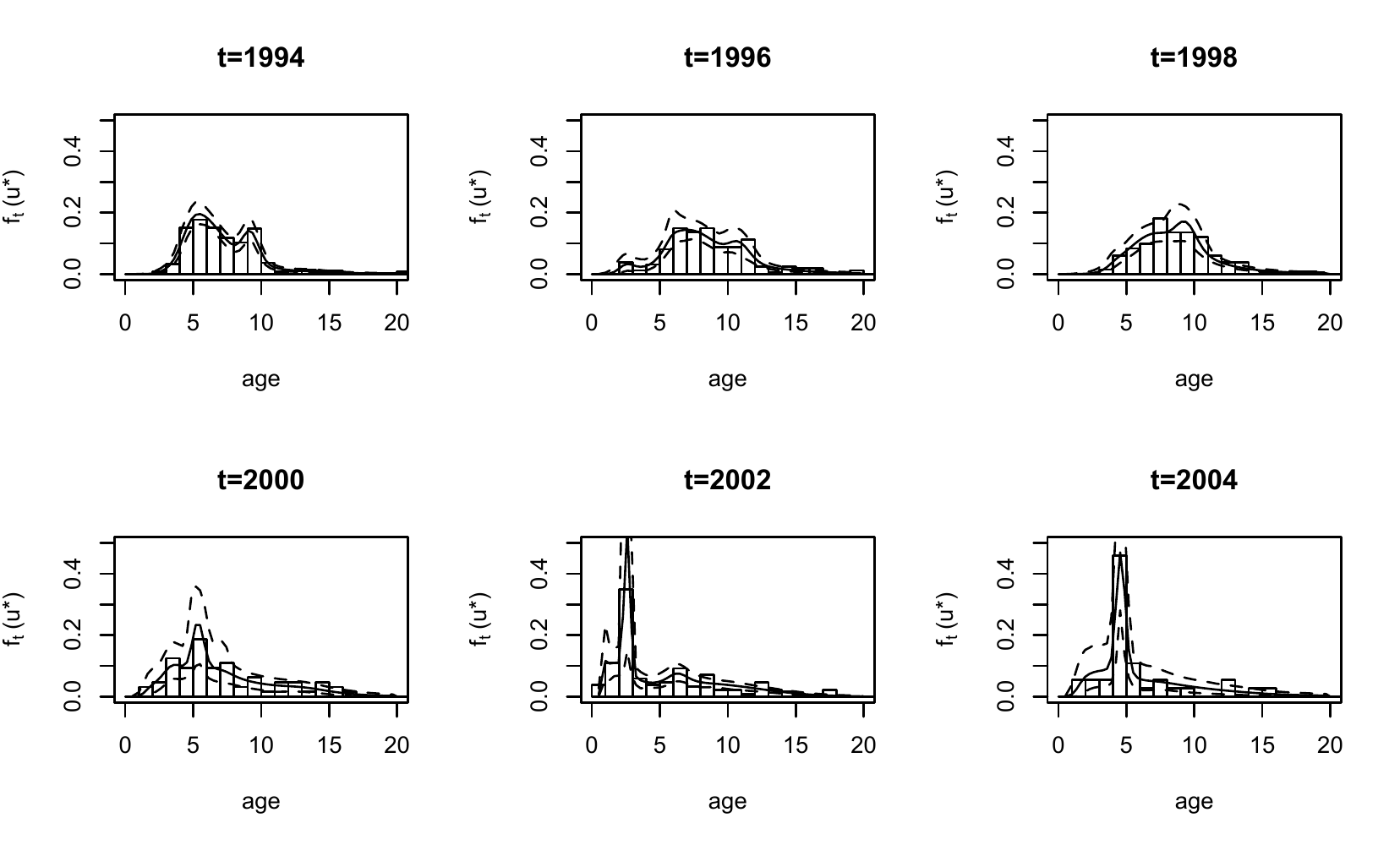}
\caption{Posterior mean and $95\%$ interval estimates for the density of age on a 
continuous scale across six years, with the data shown as a histogram.}
\label{fig:dens_age}
\end{figure}

We first discuss inference results for quantities not involving maturity. As illustrated with results for 
six years in Figure \ref{fig:dens_length}, the estimates for the density of length display a range 
of shapes. The interval estimates reflect the different sample sizes in these years; for instance,
in $1994$ there are $271$ observations, whereas in $2000$ and $2004$ there are only $64$ 
and $37$ observations, respectively. A feature of our modeling approach is that inference for 
the density of age can be obtained over a continuous scale. 
The corresponding estimates are shown in Figure \ref{fig:dens_age} for the same years as length. 

\begin{figure}
\centering
\includegraphics[height=8in,width=5in]{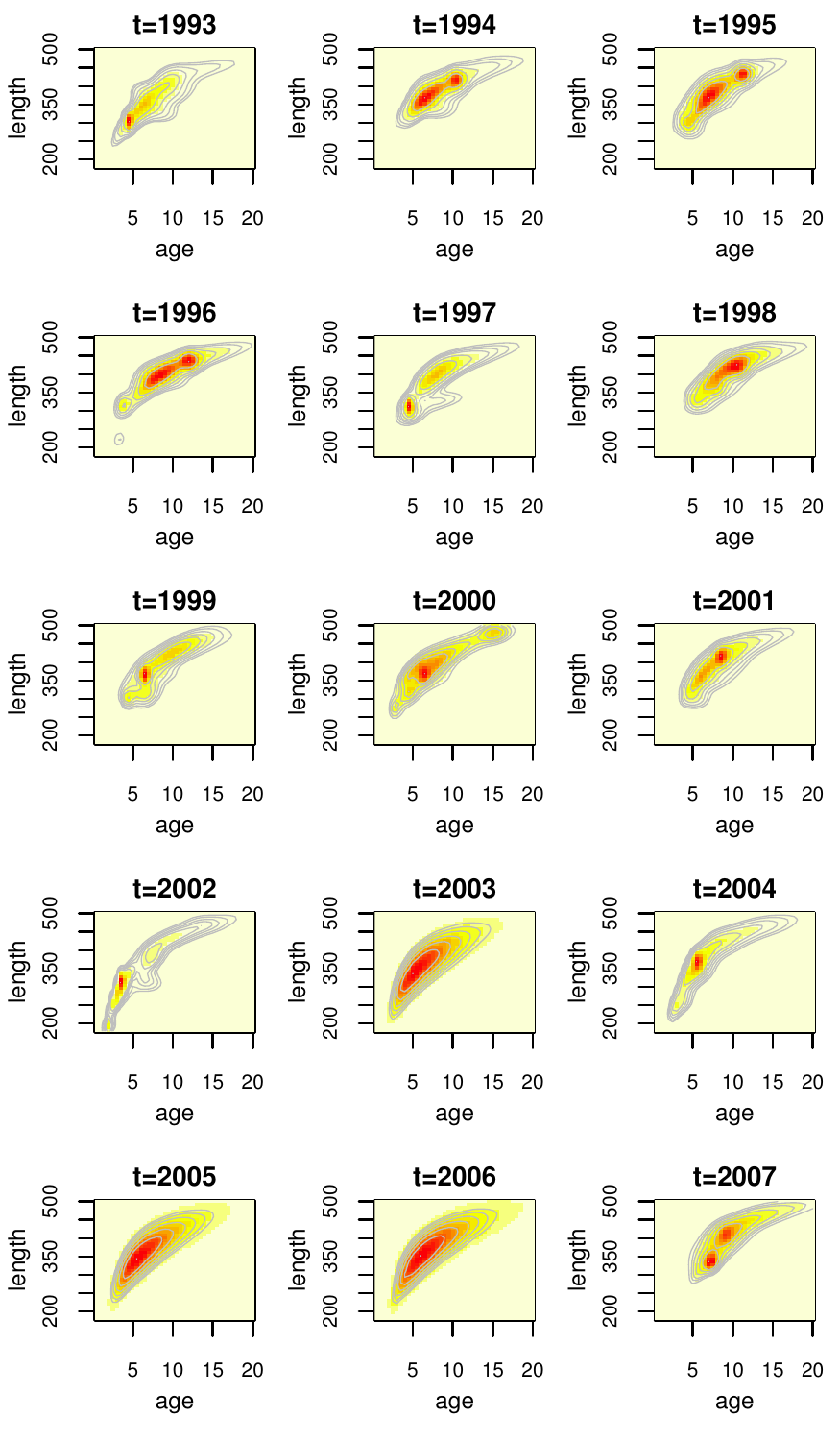}
\caption{Posterior mean estimates for the bivariate density of age and length 
across all years.}
\label{fig:dens_diffscale}
\end{figure}

\begin{figure}
\centering
\includegraphics[height=2.3in,width=5.8in]{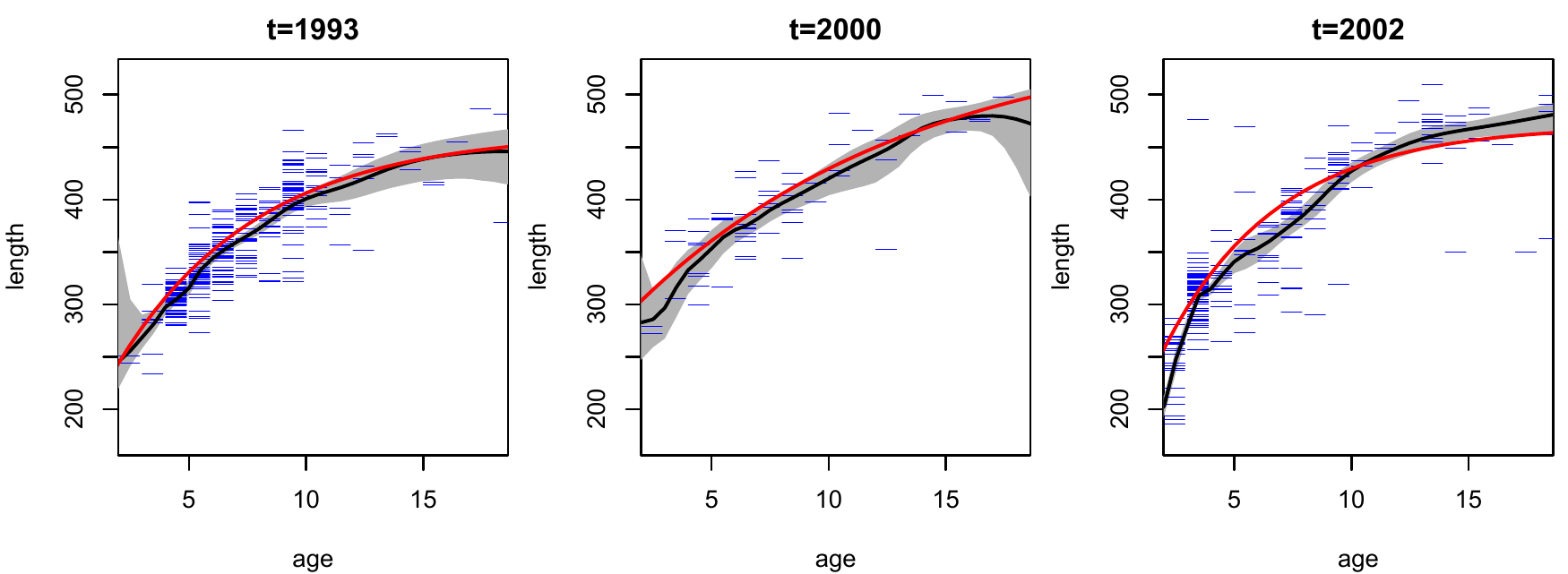}
\caption{Posterior mean and $95\%$ interval bands for the expected value of length over 
(continuous) age, $\text{E}(X\mid U^*=u^*;G_t)$, across three years. Overlaid are the 
data (in blue) and the estimated von Bertalanffy growth curves (in red).}
\label{fig:growth_curves}
\end{figure}

The posterior mean surfaces for the bivariate density of age and length are shown in Figure~\ref{fig:dens_diffscale} 
for all years, including the ones (years $2003$, $2005$ and $2006$) for which data is not available. 
The model yields more smooth shapes for the density estimates in these years.
An ellipse with a slight ``banana'' shape appears at each year, though some nonstandard features 
and differences across years are present. In particular, the density in year $2002$ extends down 
farther to smaller ages and lengths; this year is unique in that it contains a very large proportion 
of the young fish which are present in the data.
One can envision a curve going through the center of these 
densities, representing $\text{E}(X\mid U^*=u^*;G_t)$, for which we show posterior mean and $95\%$ 
interval bands for three years in Figure \ref{fig:growth_curves}. The estimates from our model are 
compared with the von Bertalanffy growth curves for length-at-age, which are based on a particular 
function of age and three parameters (estimated here using nonlinear least squares). It is noteworthy 
that the nonparametric mixture model for the joint distribution of length and age yields estimated 
growth curves which are overall similar to the von Bertalanffy parametric model, with some local
differences especially in year 2002. The uncertainty quantification in the growth curves afforded 
by the nonparametric model is important, since the attainment of unique growth curves by group 
(i.e. by location or cohort) is often used to suggest that the groups differ in some way, and this 
type of analysis should clearly take into account the uncertainty in the estimated curves.

The last year $2007$, in addition to containing few observations, is peculiar. There are no fish that 
are younger than age $6$ in this year, and most of the age $6$ and $7$ fish are recorded as immature, 
even though in all years combined, less than $10\%$ of age $6$ as well as age $7$ fish are immature. 
This year appears to be an anomaly. As there are no observations in $2005$ or $2006$, and a small 
number of observations in $2007$ which seem to contradict the other years of data, hereinafter, we 
report inferences only up to $2004$.

\begin{figure}
\centering
\includegraphics[height=7in,width=5in]{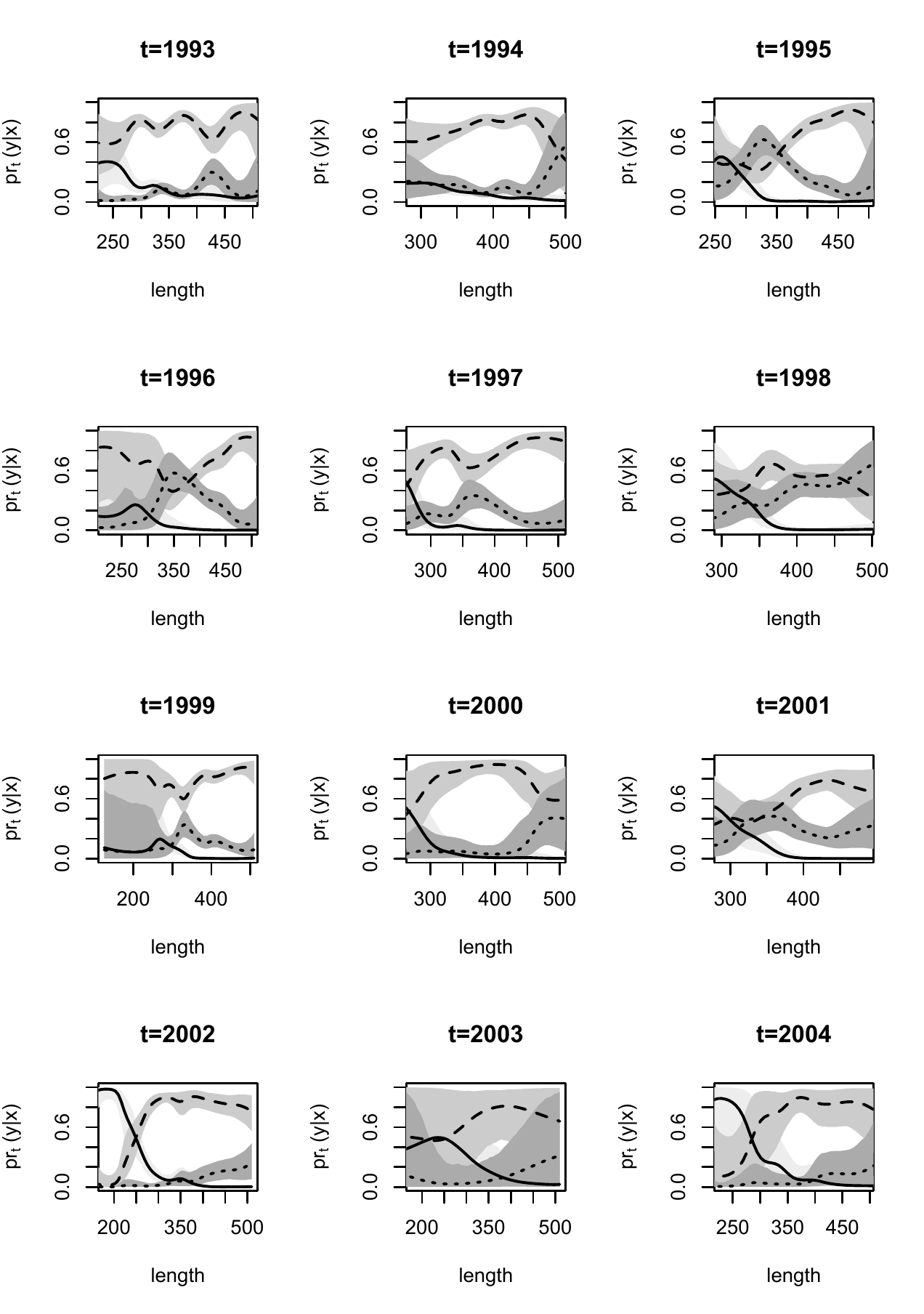}
\caption{Posterior mean (black lines) and $95\%$ interval estimates (gray shaded regions) 
for the marginal ordinal probability curves associated with length. Category 1 (immature) given by 
solid line, category 2 (pre-spawning mature) given by dashed line, and category 3 (post-spawning mature) 
shown as a dotted line.}
\label{fig:probs_ord_len_12_times}
\end{figure}

\begin{figure}
\centering
\includegraphics[height=7in,width=5in]{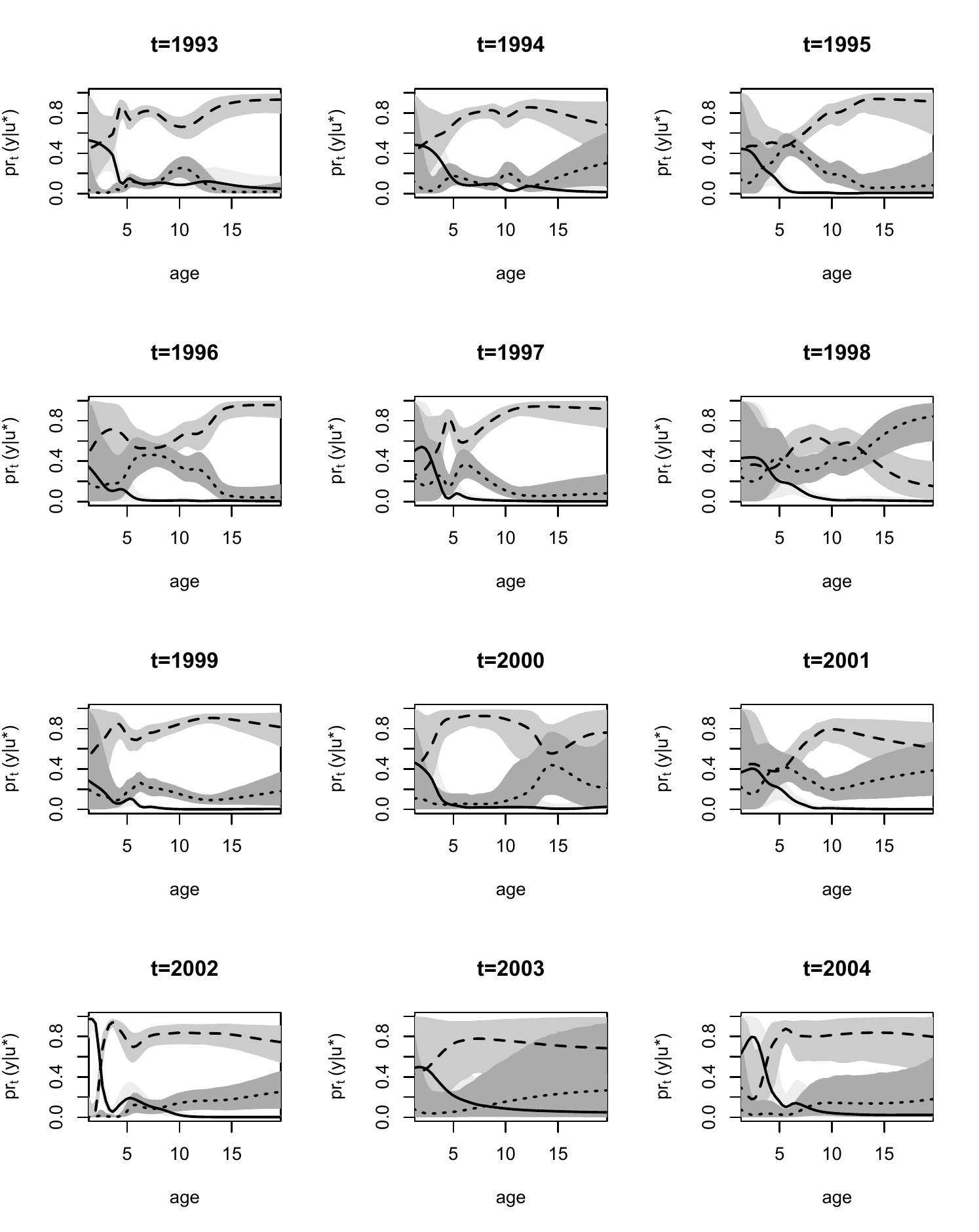}
\caption{Posterior mean (black lines) and $95\%$ interval estimates (gray shaded regions) for 
the marginal ordinal probability curves associated with age. Category 1 (immature) given by solid 
line, category 2 (pre-spawning mature) given by dashed line, and category 3 (post-spawning mature) 
shown as a dotted line.}
\label{fig:probs_ord_age_12_times}
\end{figure}

Inference for the maturation probability curves is shown over length and age in 
Figures \ref{fig:probs_ord_len_12_times} and \ref{fig:probs_ord_age_12_times}. The probability 
that a fish is immature is generally decreasing over length, reaching a value near $0$ at around 
$350$ mm in most years. There is a large change in this probability over length in $2002$ and $2004$ 
as compared to other years, as these years suggest a probability close to $1$ for very small fish near 
$200$ to $250$ mm. Turning to age, the probability of immaturity is also decreasing with age, also 
showing differences in $2002$ and $2004$ in comparison to other years. There is no clear indication 
of a general trend in the probabilities associated with levels $2$ or $3$. Years $1995$-$1997$ and 
$1999$ display similar behavior, with a peak in probability of post-spawning mature for moderate length 
values near $350$ mm, and ages $6$-$7$, favoring pre-spawning mature fish at other lengths and ages. 
The last four years $2001$-$2004$ suggest the probability of pre-spawning mature to be increasing with length up to a point and then leveling off, while post-spawning is favored most for large fish. Post-spawning appears to have a lower probability than pre-spawning mature for any age at all years, with the exception of $1998$, for which the probability associated with post-spawning is very high for older fish. 

\begin{figure}
\centering
\includegraphics[height=2.5in,width=2.8in]{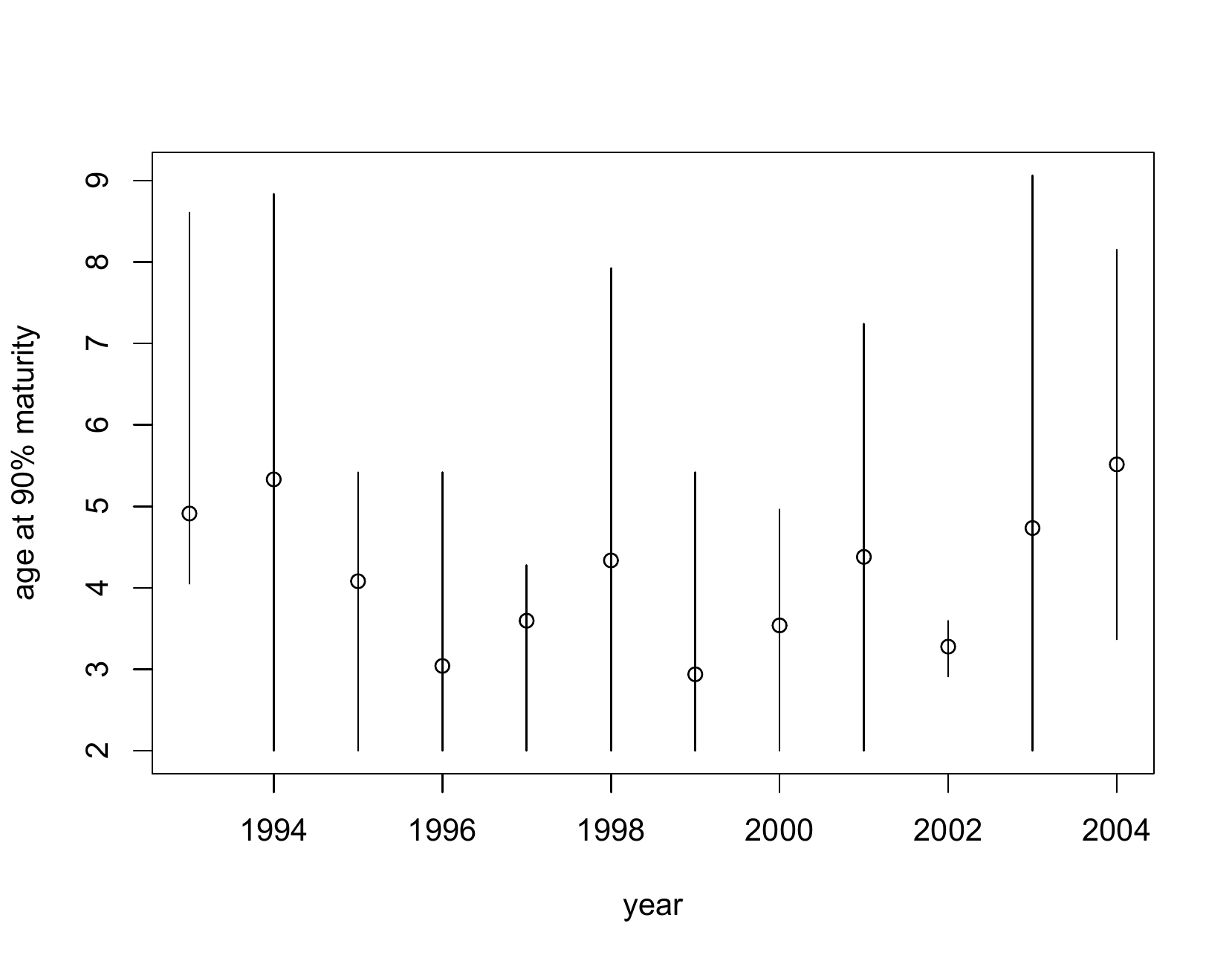}
\includegraphics[height=2.45in,width=2.8in]{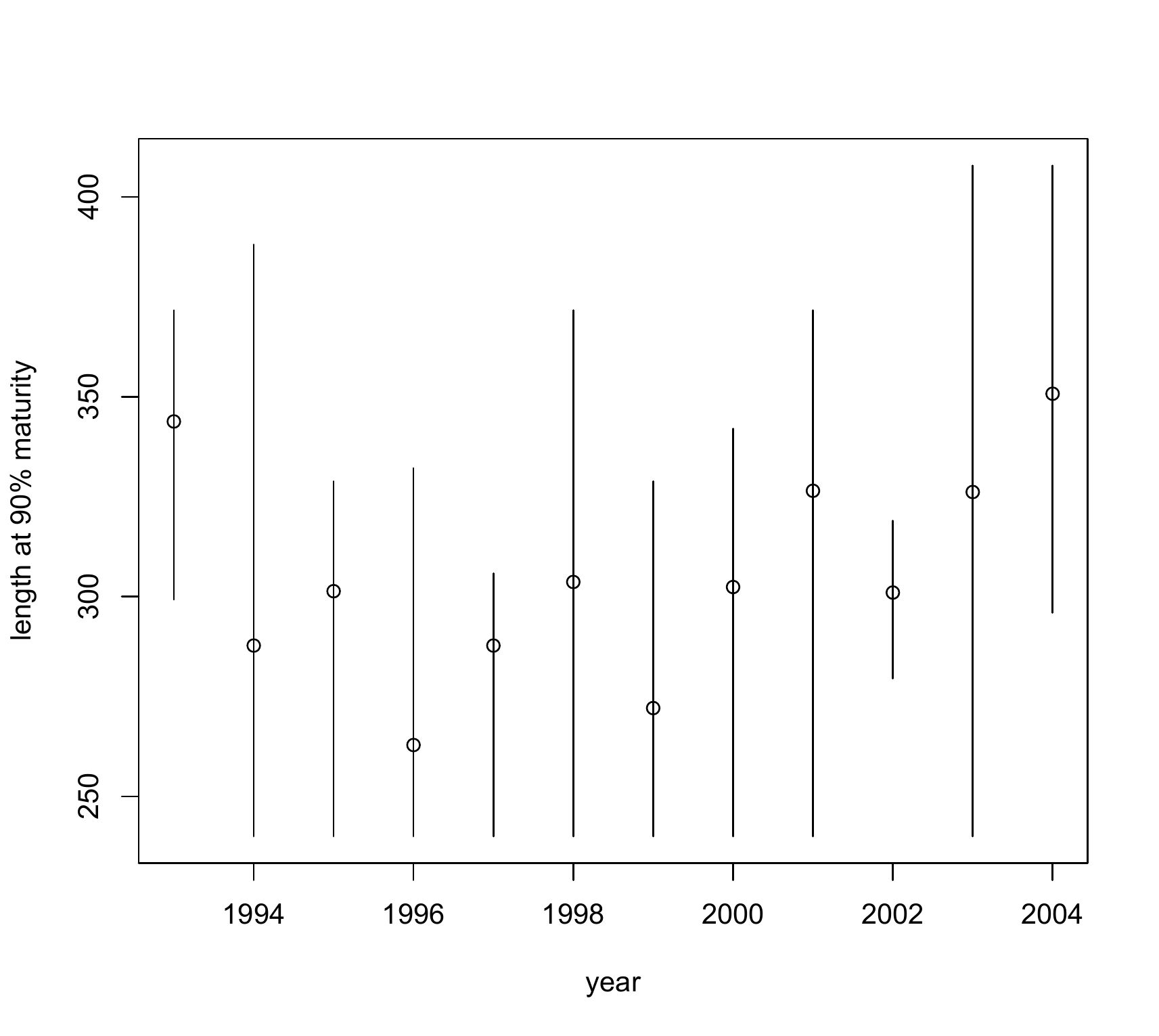}
\caption{Posterior mean and $90\%$ intervals for the smallest value of age above $2$ years at which probability of maturity first exceeds $90\%$ (left), and similar inference for length (right).}
\label{fig:age_90_mat}
\end{figure}

The Pacific States Marine Fisheries Commission states that all Chilipepper rockfish are mature at around $4$-$5$ years, and at size $304$ to $330$ mm. A stock assessment produced by the Pacific Fishery Management Council \citep{field} fitted a logistic regression to model maturity over length, from which it appears that $90\%$ of fish are mature around $300$-$350$ mm. As our model does not enforce monotonicity on the probability of maturity across age, we obtain posterior distributions for the first age greater than $2$ (since biologically all fish under $2$ should be immature) at which the probability of maturity exceeds $90\%$, given that it exceeds $90\%$ at some point. That is, for each posterior sample we evaluate $\mathrm{Pr}(Y>1\mid u^*;G_t)$ over a grid in $u^*$ beginning at $2$, and find the smallest value of $u^*$ at which this probability exceeds $0.9$. Note that there were very few posterior samples for which this probability 
did not exceed $0.9$ for any age (only $4$ samples in $1993$ and $8$ in $2003$). 
The estimates for age at $90\%$ maturity are shown in the left panel of Figure \ref{fig:age_90_mat}. 
The model uncovers a (weak) U-shaped trend across years. Also noteworthy are the very narrow interval 
bands in $2002$. Recall that this year contains an abnormally large number of young fish. In this year, 
over half of fish age $2$ (that is, of age $2$-$3$) are immature, and over $90\%$ of age $3$ (that is, of age $3$-$4$) 
fish are mature, so we would expect the age at $90\%$ maturity to be above $3$ but less than $4$, which our
estimate confirms. A similar analysis is performed for length (right panel of Figure \ref{fig:age_90_mat}) 
suggesting a trend over time which is consistent with the age analysis.

\begin{figure}[t!]
\centering
\includegraphics[height=7.5in,width=5.5in]{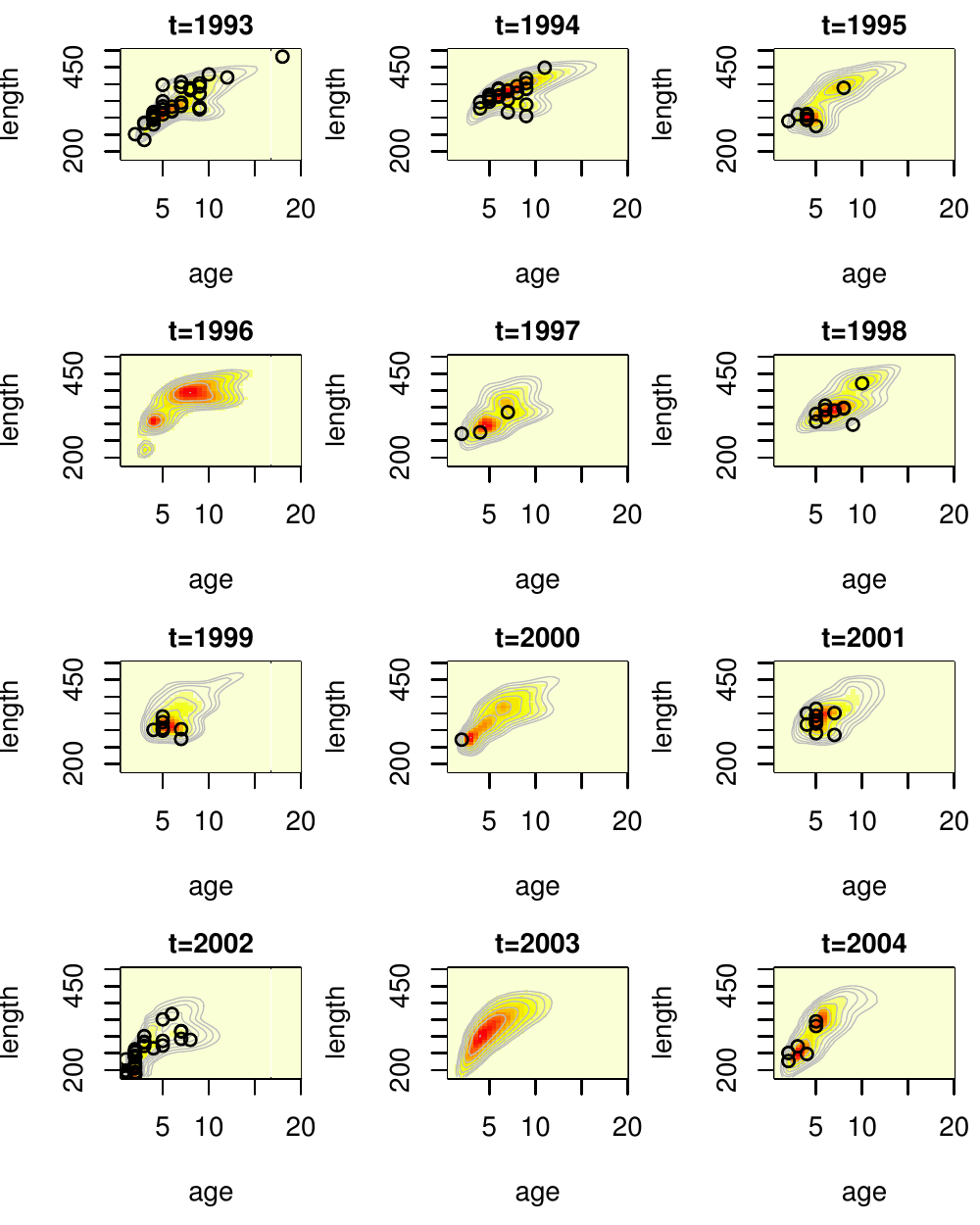}
\caption{Posterior mean estimate for the bivariate density of age and length for immature fish across years. 
The data on age and length of immature fish is overlaid in each plot.}
\label{fig:age_len_imm}
\end{figure}

Due to the monotonicity in the maturity probability curve in standard approaches, and the fact that age and length 
are treated as fixed, the point at which maturity exceeds a certain probability is a reasonable quantity to obtain in 
order to study the age or length at which most fish are mature. However, since we are modeling age and length, 
we can obtain their entire distribution at a given maturity level. These are inverse inferences, in which we study, 
for instance, $f(x \mid Y=1;G_t)$ as opposed to $\mathrm{Pr}(Y=1 \mid x;G_t)$. It is most informative to look 
at age and length for immature fish, as this makes it clear at which age or length there is essentially no probability 
assigned to the immature category. The posterior mean estimates for $f(u^*,x \mid Y=1;G_t)$ are shown 
in Figure \ref{fig:age_len_imm}. 

\subsection{Model Checking}

\begin{figure}
\centering
\includegraphics[height=3in,width=2.8in]{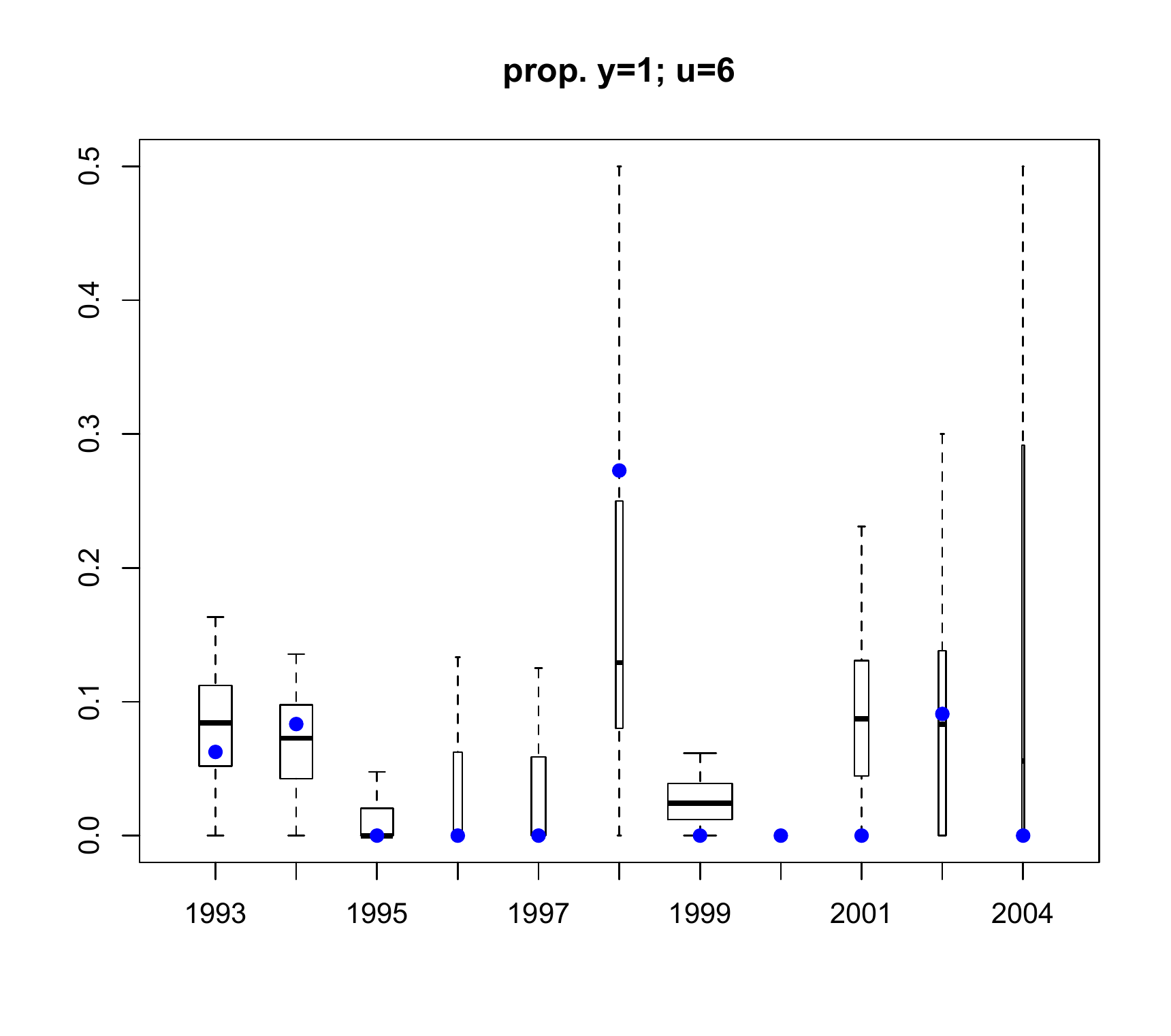}
\includegraphics[height=3in,width=2.8in]{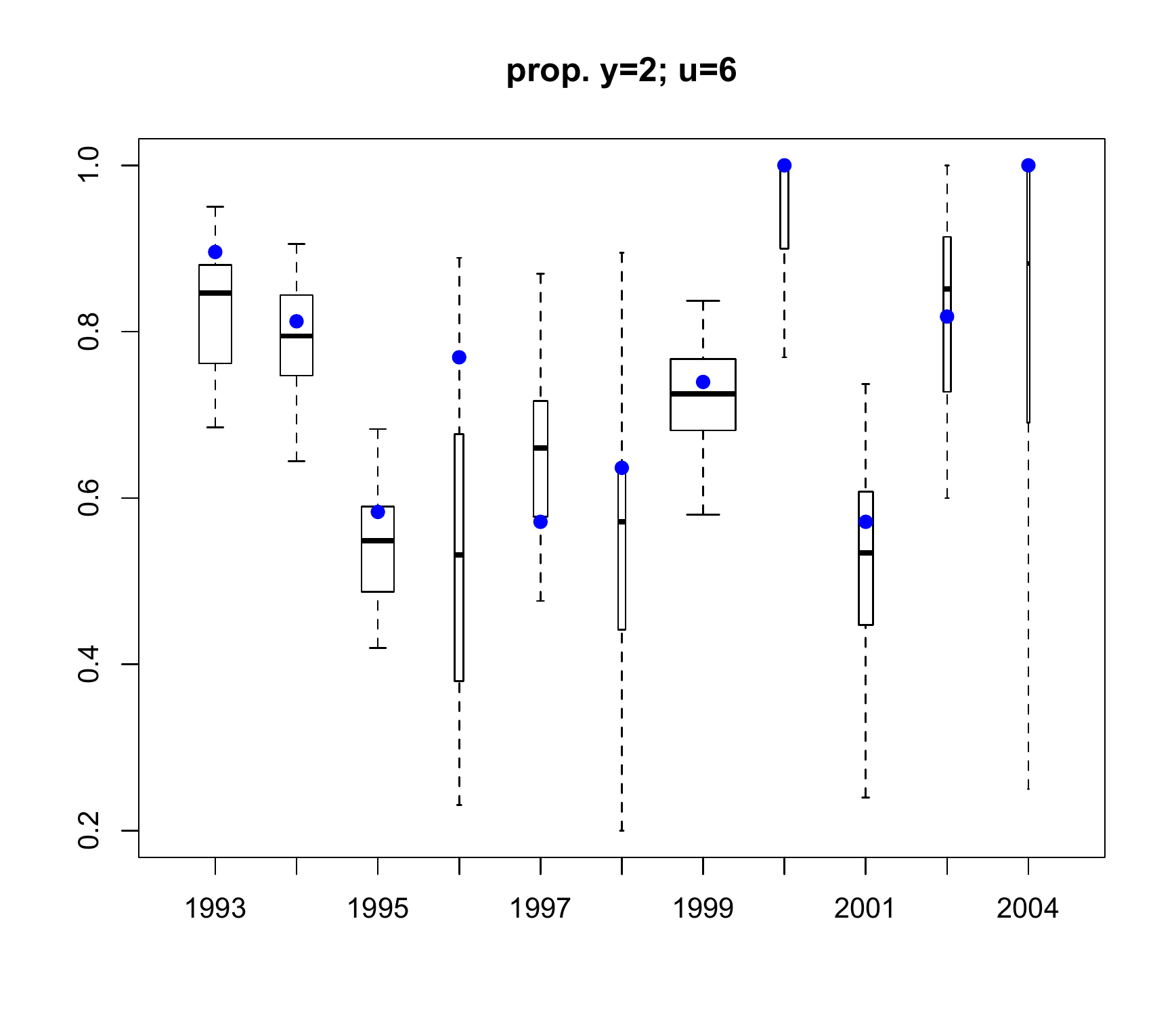}
\caption{Distributions of the proportion of age $6$ fish that were of maturity level 1 (left panel) 
and 2 (right panel) in the replicated data sets are shown as boxplots, with width proportional 
to the number of age $6$ fish in each year. The actual proportion from the data is given as a 
blue circle.}
\label{fig:pred_age6}
\end{figure}

\begin{figure}
\centering
\includegraphics[height=3in,width=2.8in]{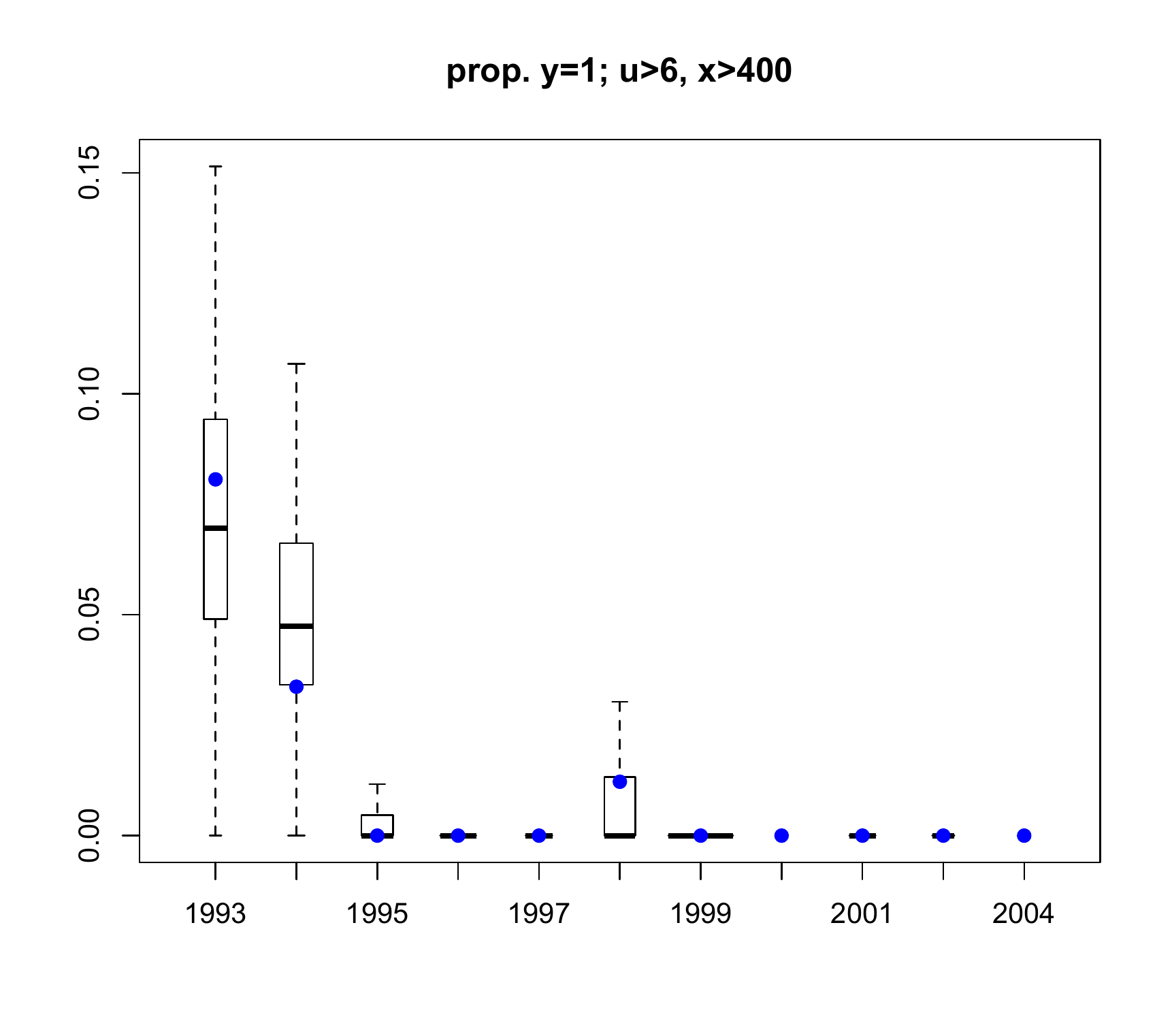}
\includegraphics[height=3in,width=2.8in]{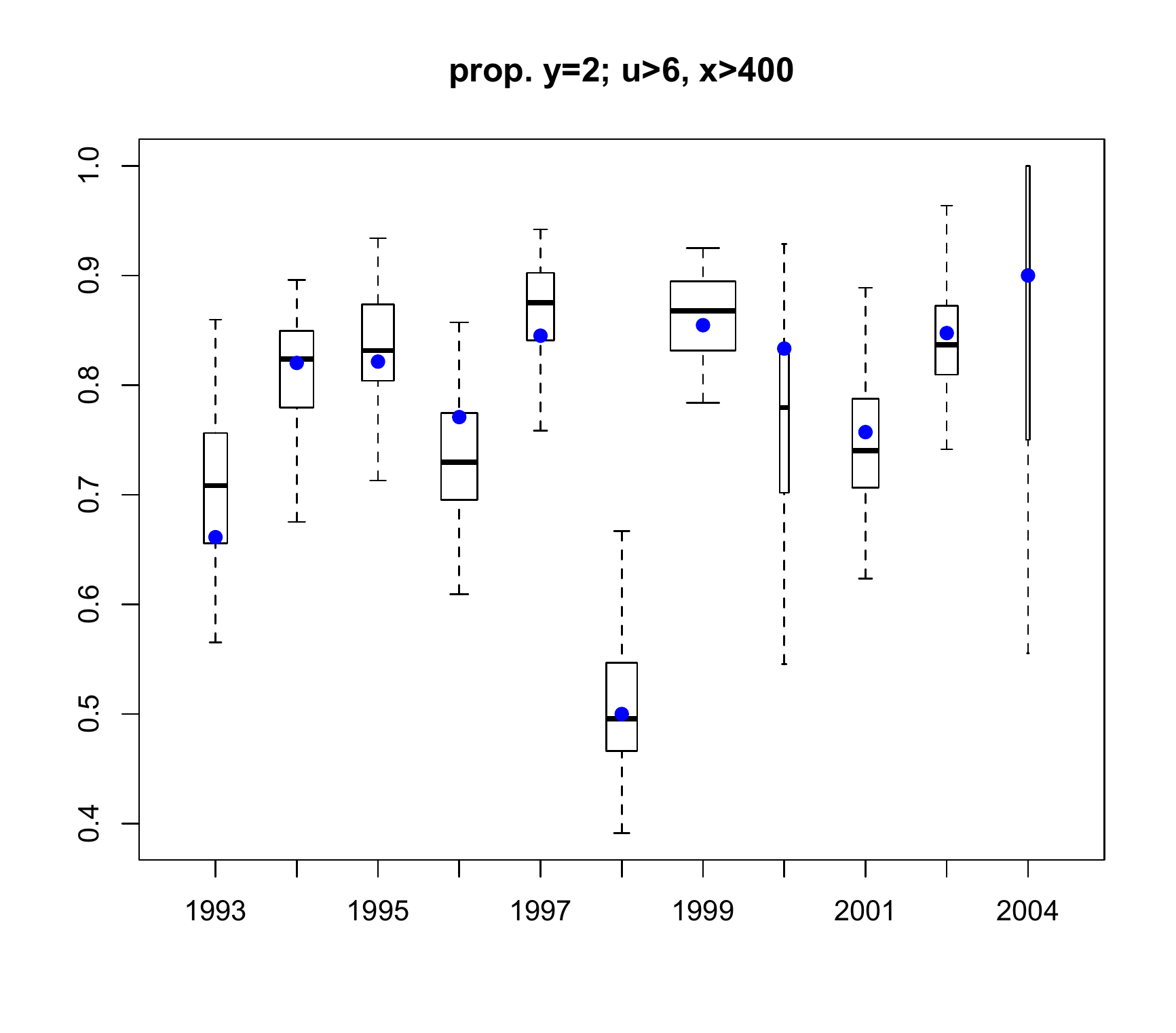}
\caption{Distributions of the proportion of fish age $7$ and above and length larger than $400$ mm 
that were of maturity level 1 (left panel) and 2 (right panel) in the replicated data sets are shown as 
boxplots, with width proportional to the number fish of this age and length in each year. The actual
proportion from the data is given as a blue circle.}
\label{fig:pred_more}
\end{figure}

\begin{figure}
\centering
\includegraphics[height=3in,width=3in]{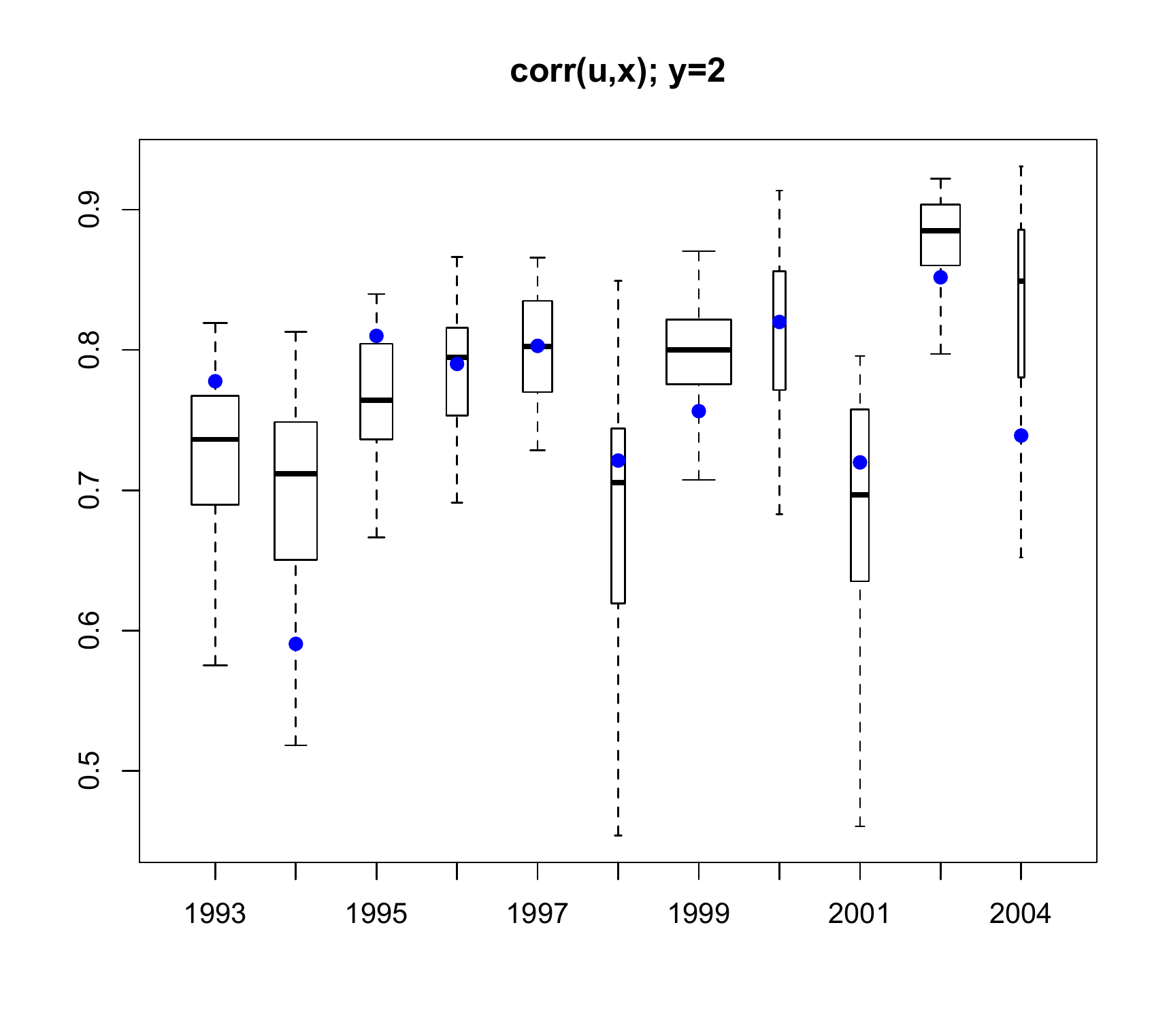}

\caption{Distributions of the sample correlation between length and age for fish that were of 
maturity level 2 in the replicated data sets are shown as boxplots, with width proportional to 
the number of level 2 fish in each year. The sample correlation based on the data is given as 
a blue circle.}
\label{fig:pred_corr}
\end{figure}

Here, we discuss results from posterior predictive model checking. In particular, we generated
replicate data sets from the posterior predictive distribution, and compared to the real data 
using specific test quantities \citep{gcsr}. Using the MCMC output, we simulate replicate 
data sets $\{(y_{ti},w_{ti},x_{ti})^{rep}\}$ of the same size as the original data. We then choose 
some test quantity, $T(\{y_{ti},w_{ti},x_{ti}\})$, and for each replicate data set at time $t$, 
determine the value of the test quantity and compare the distribution of test quantities 
with the value computed from the actual data. To obtain Figure \ref{fig:pred_age6} we 
computed, for each replicate sample, the proportion of age $6$ fish that were of maturity 
levels $1$ and $2$. Boxplots of these proportions are shown, with the actual proportions 
from the data indicated as blue points. The width of each box is proportional to the number 
of age $6$ fish in that year. Figure \ref{fig:pred_more} refers to fish of at least age $7$ 
with length larger than $400$ mm. Finally, Figure \ref{fig:pred_corr} plots the sample 
correlation of length and age for fish of maturity level $2$. The results reported in 
Figures \ref{fig:pred_age6}, \ref{fig:pred_more} and \ref{fig:pred_corr} suggest that the 
model is predicting data which is very similar to the observed data in terms of practically 
important inferences.

Although results are not shown here, we also studied residuals with cross-validation, 
randomly selecting $20\%$ of the observations in each year and refitting the model, 
leaving out these observations. 
We obtained residuals $\tilde{y}_{ti}-\text{E}({Y} \mid W=\tilde{w}_{ti},X=\tilde{x}_{ti};G_t)$ 
for each observation $(\tilde{y}_{ti},\tilde{w}_{ti},\tilde{x}_{ti})$ which was left out. There 
was no apparent trend in the residuals across covariate values, that is, no indication 
that we are systematically under or overestimating fish maturity of a particular length 
and/or age. 

\section{Discussion}
\label{sec:disc}

The methods developed for dynamic ordinal regression are widely applicable to modeling mixed ordinal-continuous distributions indexed in discrete time. At any particular point in time, the DP mixture representation for the latent response-covariate distribution is retained, enabling flexible inference for a variety of functionals, and allowing 
standard posterior simulation techniques for DP mixture models to be utilized. 

In contrast to standard approaches to ordinal regression, the model does not force specific trends, such as monotonicity, 
in the regression functions. We view this as an attribute in most settings. Nevertheless, in situations in which it is believed that monotonicity exists, we must realize that the data will determine the model output, and may not produce strictly monotonic relationships. 
In the fish maturity example, it is generally accepted that monotonicity exists in the relationship between maturity and age or length. Although our model does not enforce this, the inferences generally agree with what is expected to be true biologically. Our model is also extremely relevant to this setting, as the covariates age and length are treated as random, and the ordinal nature of recorded age is accounted for using variables which represent underlying continuous age. The set of inferences that are provided under this framework, including estimates for length as a function of age, make this modeling approach powerful for the particular application considered, as well as related problems.

While year of sampling was considered to be the index of dependence in this analysis, an alternative is to consider cohort as an index of dependence. All fish born in the same year, or the same age in a given year, represent one cohort. Grouping fish by cohort rather than year of record should lead to more homogeneity within a group, however there are also some possible issues since fish will generally be younger as cohort index increases. This is a consequence of having a particular set of years for which data is collected, i.e., the cohort of fish born in $2006$ can not be older than $4$ if data collection stopped in $2009$. Due to complications such as these, combined with exploration of the relationships within each cohort, we decided to treat year of data collection as the index of dependence, but cohort indexing could be more appropriate in other
analyses of similar data structures.
 
The proposed modeling approach could also be useful in applications in finance. One such example arises in the analysis of price changes of stocks. In the past, stocks traded on the New York Stock Exchange were priced in eighths, later moved to sixteenths, and corporate bonds still trade in eighths. In analyzing price changes of stocks which are traded in fractions, it is inappropriate to treat the measurements as continuous, particularly if the range of values is not very large \citep[e.g.,][]{mullerczado}. The price changes should be treated with a discrete response model, and the possible responses are ordered, ranging from a large negative return to a large positive return. One possible analysis may involve modeling the monthly returns as a function of covariates such as trade volume, and must take into account the ordinal nature of the responses. In addition, the distribution of returns in a particular month is likely correlated with the previous month, and the regression relationships must be allowed to be related from one month to the next. In finance as well as environmental science, empirical distributions may exhibit non-standard features which require more general methods,
such as the nonparametric mixture model developed here.

%
%

\vspace{1cm}

\singlespacing
\bibliographystyle{asa}
\bibliography{ddp_paper_bib}

\appendix
\section{Properties of the DDP Prior Model}\label{sec:app_corr}
Here, we provide derivations of the various correlations associated with the DDP prior, 
as given in Section \ref{sec:new_ddp}.

\subsection*{Autocorrelation of $(\beta_t,\beta_{t+k})$}
Since the process is stationary with $\beta_t\sim \text{beta}(\alpha,1)$ at any time $t$, 
$\text{E}(\beta_t \mid \alpha)=$ $\alpha/(\alpha+1)$ and $\text{var}(\beta_t \mid \alpha)=$
$\alpha/\{ (\alpha+1)^2(\alpha+2) \}$.
We also have 
\begin{equation}
\text{E}(\beta_t\beta_{t+k} \mid \alpha,\phi) =
\text{E}\left\{ \exp({-\zeta^2/\alpha}) \right\} 
\text{E}\left\{ \exp( - (\eta_t^2+\eta_{t+k}^2) / 2\alpha) \right\}
\tag{A.1}
\end{equation}
using the definition of the $\mathcal{B}$ process in (\ref{eqn:beta_sp}). The first expectation 
can be obtained through the moment generating function of $\zeta^2 \sim \chi_1^2$, which is 
given by $\text{E}(e^{t \zeta^{2}})=$ $(1-2t)^{-1/2}$, for $t<1/2$. Hence, for $t=-1/\alpha$, we obtain 
$\text{E}\left\{ \exp({-\zeta^2/\alpha}) \right\} =$ $\alpha^{1/2}/(2+\alpha)^{1/2}$. Regarding the 
second expectation, note that $(\eta_t,\eta_{t+k}) \sim \mathrm{N}(0,C_k)$, with $C_k$ a covariance 
matrix with diagonal elements equal to 1 and off-diagonal elements equal to $\rho_k$. 
Integration results in $\text{E}\left\{ \exp( - (\eta_t^2+\eta_{t+k}^2)/2\alpha) \right\}=$
$\alpha (1-\rho_k^2)^{1/2}/\{ (1-\rho_k^2+\alpha)^2-\alpha^2\rho_k^2 \}^{1/2}$.
The correlation in (\ref{eqn:acf_beta}) results by combining the terms above with the 
expressions for $\text{E}(\beta_{t} \mid \alpha)$ and $\text{var}(\beta_{t} \mid \alpha)$.

\subsection*{Autocorrelation of DP weights}

First, $\text{E}(p_{l,t} \mid \alpha)=$
$\text{E}\{(1-\beta_{l,t})\prod_{r=1}^{l-1}\beta_{r,t} \mid \alpha \}$. Since the $\beta_{l,t}$ are 
independent across $l$, and $\text{E}(\beta_{l,t} \mid \alpha)=$ $\alpha/(\alpha+1)$,
we obtain $\text{E}(p_{l,t} \mid \alpha)=$ $\alpha^{l-1}/(1+\alpha)^{l}$. Similarly, 
$\text{E}(p_{l,t}^{2} \mid \alpha)=$ 
$\text{E}\{ (1-\beta_{l,t})^{2} \mid \alpha \} \prod_{r=1}^{l-1} \text{E}(\beta_{r,t}^{2} \mid \alpha)=$
$2 \alpha^{l-1}/\{ (\alpha+1)(\alpha + 2)^{l} \}$, from which $\text{var}(p_{l,t} \mid \alpha)$ 
obtains.

Since $p_{l,t} p_{l,t+1} =$ $(1-\beta_{l,t})(1-\beta_{l,t+1})\prod_{r=1}^{l-1}\beta_{r,t}\beta_{r,t+1}$,
and $(\beta_{l,t},\beta_{l,t+1})$ is independent of $(\beta_{m,t},\beta_{m,t+1})$, for any $l \neq m$,
we can write 
\[
\text{E}(p_{l,t}p_{l,t+1}\mid \alpha, \phi) = \text{E}\{ (1-\beta_{l,t})(1-\beta_{l,t+1}) \mid \alpha,\phi \} 
\prod_{r=1}^{l-1} \text{E}( \beta_{r,t} \beta_{r,t+1} \mid \alpha,\phi ).
\]
The required expectations in the above equation can be obtained from (A.1) for $k=1$, such 
that $\rho_{1} = \phi$. Combining the above expressions yields the covariance in (\ref{eqn:covw}).

\subsection*{Autocorrelation of consecutive distributions}

The DDP prior model implies at any $t$ a $\text{DP}(\alpha,G_0)$ prior for 
$G_t=$ $\sum_{l=1}^\infty p_{l,t} \delta_{\boldsymbol{\theta}_l}$, where the $\boldsymbol{\theta}_l$ 
are i.i.d. from a distribution $G_{0}$ on $\mathbb{R}^M$. Hence, for any measurable subset 
$A \subset \mathbb{R}^{M}$, we have $\text{E}(G_t(A) \mid \alpha,G_0)=$ 
$\text{E}(G_{t+1}(A)\mid \alpha,G_0) = G_0(A)$, and $\text{var}(G_t(A) \mid \alpha,G_0)=$
$\text{var}(G_{t+1}(A)\mid \alpha,G_0)=$ $G_0(A)(1-G_0(A))/(\alpha+1)$.

The additional expectation needed in order to obtain $\text{corr}(G_t(A),G_{t+1}(A) \mid \phi,\alpha,G_0)$
is $\text{E}(G_t(A)G_{t+1}(A)\mid \phi,\alpha,G_0) =$ 
$\text{E}\left\{(\sum\nolimits_{l=1}^{\infty}p_{l,t}\delta_{\boldsymbol{\theta}_l}(A))
(\sum\nolimits_{m=1}^{\infty}p_{m,t+1}\delta_{\boldsymbol{\theta}_m}(A))\mid \phi,\alpha,G_0\right\}$, 
which can be written as
\begin{equation}
\text{E}\left(\sum\nolimits_{l=1}^{\infty}p_{l,t}p_{l,t+1}(\delta_{\boldsymbol{\theta}_l}(A))^2 \right) +
\text{E}\left(\sum\nolimits_{l=1}^\infty \sum\nolimits_{m \neq l}p_{l,t}p_{m,t+1}\delta_{\boldsymbol{\theta}_l}(A)
\delta_{\boldsymbol{\theta}_m}(A) \right).
\tag{A.2}
\end{equation}
For the first expectation in (A.2), note that $p_{l,t}p_{l,t+1}$ is independent of $(\delta_{\boldsymbol{\theta}_l}(A))^2$, 
and $\text{E}(p_{l,t}p_{l,t+1} \mid \alpha,\phi)$ has been derived earlier. Moreover, since
$(\delta_{\boldsymbol{\theta}_l}(A))^{2}$ is equal to $1$ if $\boldsymbol{\theta}_l \in A$ and $0$ otherwise, 
$\text{E}( (\delta_{\boldsymbol{\theta}_l}(A))^2 \mid \alpha,G_0)=G_0(A)$. Regarding the second expectation
in (A.2), we have again that $p_{l,t} p_{m,t+1}$ and $\delta_{\boldsymbol{\theta}_l}(A) \delta_{\boldsymbol{\theta}_m}(A)$
are independent. Here, $\text{E}(\delta_{\boldsymbol{\theta}_l}(A) \delta_{\boldsymbol{\theta}_m}(A) \mid \alpha,G_0)=$
$(G_0(A))^{2}$, since $\boldsymbol{\theta}_l$ and $\boldsymbol{\theta}_m$ are independent for $m \neq l$.
The final step of the derivation involves the expectations 
$\text{E}(p_{l,t} p_{m,t+1} \mid \alpha,\phi)$, for $m \neq l$.
Note that, if $l < m$, $p_{l,t} p_{m,t+1}=$ $(\prod_{r=1}^{l-1} \beta_{r,t}\beta_{r,t+1}) \, \beta_{l,t+1}(1-\beta_{l,t}) \, 
(\prod_{r=l+1}^{m-1} \beta_{r,t+1}) \, (1-\beta_{m,t+1})$, and an analogous expression can be written 
when $m < l$. Therefore, for general $m \neq l$, $p_{l,t} p_{m,t+1}$ can be expressed as a product of 
$\max\{l,m\}$ independent components, since each component comprises one or two of the random 
variables in $\{\beta_{l,t}\}$ and $\{\beta_{m,t+1}\}$, in the latter case having the same first subscript. 
Thus, $\text{E}(p_{l,t} p_{m,t+1} \mid \alpha,\phi)$ can be developed through products of expectations 
of the form $\text{E}(\beta_{l,t}\mid \alpha)$ and $\text{E}(\beta_{l,t}\beta_{l,t+1} \mid \alpha,\phi)$, 
which have been obtained earlier.

\section{Posterior Simulation Details}
\label{sec:app_mcmc}

We derive the posterior full conditionals and provide updating strategies for many of the parameters of the hierarchical model in Section \ref{sec:fish_methods}.

\subsubsection*{Updating the weights}

 The full conditional for $(\{\zeta_l\},\{\eta_{l,t}\})$ is given by $p(\{
\zeta_l\},\{\eta_{l,t}\}\mid \dots,\text{data})\propto$
\[ \prod_{l=1}^{N-1}\text{N}(\zeta_l;0,1)\text{N}(\eta_{l,1};0,1)\prod_{t=2}^T\prod_{l=1}^{N-1}\text{N}(\eta_{l,t};\phi\eta_{l,t-1},1-\phi^2)\prod_{t=1}^T\prod_{i=1}^{n_t}\sum_{l=1}^Np_{l,t}\delta_l(L_{t,i}).\]
Write $\prod_{i=1}^{n_t}\sum_{l=1}^Np_{l,t}\delta_l(L_{t,i})=\prod_{l=1}^Np_{l,t}^{M_{l,t}}$, where $M_{l,t}=\mid \{(t,i):L_{t,i}=l\}\mid $, i.e., the number of observations at time $t$ assigned to component $l$. Filling in the form for $\{p_{l,t}\}$ gives
 \begin{multline*}\prod_{i=1}^{n_t}\sum_{l=1}^Np_{l,t}\delta_l(L_{t,i})=\left(1-\exp\left(-\frac{\zeta_1^2+\eta_{1,t}^2}{2\alpha}\right)\right)^{M_{1,t}}\exp\left(-\frac{M_{N,t}\sum_{l=1}^{N-1}(\zeta_l^2+\eta_{l,t}^2)}{2\alpha}\right)\\
\prod_{l=2}^{N-1}\left\{\left(1-\exp\left(-\frac{\zeta_l^2+\eta_{l,t}^2}{2\alpha}\right)\right)^{M_{l,t}}\exp\left(-\frac{M_{l,t}\sum_{r=1}^{l-1}(\zeta_r^2+\eta_{r,t}^2)}{2\alpha}\right)\right\}.\end{multline*}

The full conditional for each $\zeta_l$, $l=1,\dots,N-1$, is therefore
\[p(\zeta_l\mid \dots,\text{data})\propto \exp\left(-\frac{\zeta_l^2}{2}\right)\exp\left(\frac{-\zeta_l^2\sum_{t=1}^T\sum_{r=l+1}^NM_{r,t}}{2\alpha}\right)\prod_{t=1}^T\left(1-\exp\left(-\frac{\zeta_l^2+\eta_{l,t}^2}{2\alpha}\right)\right)^{M_{l,t}}\]
giving
\[p(\zeta_l\mid \dots,\text{data})\propto \text{N}(\zeta_l;0,(1+\alpha^{-1}\sum_{t=1}^T\sum_{r=l+1}^NM_{r,t})^{-1})\prod_{t=1}^T\left(1-\exp\left(-\frac{\zeta_l^2+\eta_{l,t}^2}{2\alpha}\right)\right)^{M_{l,t}}\]
We use a slice sampler to update $\zeta_l$, with the following steps:
\begin{itemize}
\item Draw $u_t\sim \text{uniform}\left(0,\left(1-\exp\left(-\frac{\zeta_l^2+\eta_{l,t}^2}{2\alpha}\right)\right)^{M_{l,t}}\right)$, for $t=1,\dots,T.$
\item Draw $\zeta_l\sim \text{N}(0,(1+\alpha^{-1}\sum_{t=1}^T\sum_{r=l+1}^NM_{r,t})^{-1})$, restricted to the lie in the interval
$\left\{\zeta_l:u_t<\left(1-\exp\left(-\frac{\zeta_l^2+\eta_{l,t}^2}{2\alpha}\right)\right)^{M_{l,t}},\,t=1,\dots,T\right\}$. Solving for $\zeta_l$ in each of these $T$ equations gives $\zeta_l^2>-\eta_{l,t}^2-2\alpha \log(1-u_t^{1/M_{l,t}})$, for $t=1,\dots,T$. Therefore, if $-\eta_{l,t}^2-2\alpha \log(1-u_t^{1/M_{l,t}})<0$ for all $t$, then $\zeta_l$ has no restrictions, and is therefore sampled from a normal distribution. Otherwise, if $-\eta_{l,t}^2-2\alpha \log(1-u_t^{1/M_{l,t}})>0$ for some $t$, then $\mid \zeta_l\mid >\max_t\{(-\eta_{l,t}^2-2\alpha \log(1-u_t^{1/M_{l,t}}))^{1/2}\}$.  This then requires sampling $\zeta_l$ from a normal distribution, restricted to the intervals $(-\infty, -\max_t\{(-\eta_{l,t}^2-2\alpha \log(1-u_t^{1/M_{l,t}}))^{1/2}\})$, and $(\max_t\{(-\eta_{l,t}^2-2\alpha \log(1-u_t^{1/M_{l,t}}))^{1/2}\},\infty)$.
\end{itemize}
In the second step above, we may have to sample from a normal distribution, restricted to two disjoint intervals. The resulting distribution is therefore a mixture of two truncated normals, with probabilities determined by the (normalized) probability the normal assigns to each interval. These truncated normals both have mean $0$ and variance $(1+\sum_{t=1}^T\sum_{r=l+1}^NM_{r,t}/\alpha)^{-1}$, and each mixture component has equal probability.

The full conditional for each $\eta_{l,t}$, $l=1,\dots,N-1$, $t=2,\dots,T-1$, is proportional to
{\small
\[\text{N}\left(\eta_{l,t};0,\frac{\alpha}{\sum_{r=l+1}^NM_{r,t}}\right)\text{N}(\eta_{l,t};\phi\eta_{l,t-1},1-\phi^2)\text{N}(\eta_{l,t+1};\phi\eta_{l,t},1-\phi^2)
\left(1-\exp\left(-\frac{\zeta_l^2+\eta_{l,t}^2}{2\alpha}\right)\right)^{M_{l,t}}\]
\begin{multline*} \propto \text{N}\left(\eta_{l,t};\frac{\phi\alpha(\eta_{l,t-1}+\eta_{l,t+1})}{\phi^2(\alpha-\sum_{r=l+1}^NM_{r,t})+\alpha+\sum_{r=l+1}^NM_{r,t}},\frac{\alpha(1-\phi^2)}{\phi^2(\alpha-\sum_{r=l+1}^NM_{r,t})+\alpha+\sum_{r=l+1}^NM_{r,t}}\right)\\ \left(1-\exp\left(-\frac{\zeta_l^2+\eta_{l,t}^2}{2\alpha}\right)\right)^{M_{l,t}}\end{multline*}}
Each $\eta_{l,t}$, $l=1,\dots,N-1$, and $t=2,\dots,T-1$, can therefore be sampled with a slice sampler:
\begin{itemize}
\item Draw $u\sim \text{Unif}\left(0,\left(1-\exp\left(-\frac{\zeta_l^2+\eta_{l,t}^2}{2\alpha}\right)\right)^{M_{l,t}}\right)$.
\item Draw $\eta_{l,t}\sim \text{N}\left(\eta_{l,t};\frac{\phi\alpha(\eta_{l,t-1}+\eta_{l,t+1})}{\phi^2(\alpha-\sum_{r=l+1}^NM_{r,t})+\alpha+\sum_{r=l+1}^NM_{r,t}},\frac{\alpha(1-\phi^2)}{\phi^2(\alpha-\sum_{r=l+1}^NM_{r,t})+\alpha+\sum_{r=l+1}^NM_{r,t}}\right)$, restricted to $\left\{\eta_{l,t}:\left(1-\exp\left(-\frac{\zeta_l^2+\eta_{l,t}^2}{2\alpha}\right)\right)^{M_{l,t}}>u\right\}$, giving $\eta_{l,t}^2>-2\alpha \log(1-u^{1/M_{l,t}})-\zeta_l^2$.
\end{itemize}
In the second step above, we will again either sample from a single normal or a mixture of truncated normals, where each normal has the same mean and variance, but the truncation intervals differ. Since the mean of this normal is not zero, the weights assigned to each truncated normal are not the same. The unnormalized weight assigned to the normal which places positive probability on $((-2\alpha \log(1-u^{1/M_{l,t}})-
\zeta_l^2)^{1/2},\infty)$ is given by $1-F((-2\alpha \log(1-u^{1/M_{l,t}})-\zeta_l^2)^{1/2})$, where $F$ is the CDF of the normal for $\eta_{l,t}$ given in the second step. The unnormalized weight given to the component which places positive probability on $(-\infty,-(-2\alpha \log(1-u^{1/M_{l,t}})-\zeta_l^2)^{1/2})$ is given by $F(-(-2\alpha \log(1-u^{1/M_{l,t}})-\zeta_l^2)^{1/2})$.

The full conditionals for $\eta_{l,1}$ and $\eta_{l,T}$ are slightly different. The full conditional for $\eta_{l,1}$ is
{\small \[ p(\eta_{l,1}\mid \dots,\text{data})\propto \text{N}(\eta_{l,1};0,\frac{\alpha}{\sum_{r=l+1}^NM_{r,1}})\text{N}(\eta_{l,1};0,1)\text{N}(\eta_{l,2};\phi\eta_{l,1},1-\phi^2)\left(1-\exp\left(-\frac{\zeta_l^2+\eta_{l,1}^2}{2\alpha}\right)\right)^{M_{l,1}},\]}
\begin{multline*} \propto \text{N}\left(\eta_{l,1};\frac{\phi\alpha\eta_{l,2}}{\alpha+\sum_{r=l+1}^NM_{r,1}-\phi^2\sum_{r=l+1}^NM_{r,1}},\frac{\alpha(1-\phi^2)}{\alpha+\sum_{r=l+1}^NM_{r,1}-\phi^2\sum_{r=l+1}^NM_{r,1}}\right)\\
\left(1-\exp\left(-\frac{\zeta_l^2+\eta_{l,1}^2}{2\alpha}\right)\right)^{M_{l,1}}.
\end{multline*}

For $\eta_{l,T}$, we have:
\[p(\eta_{l,T}\mid \dots,\text{data}) \propto \text{N}(\eta_{l,T};0,\frac{\alpha}{\sum_{r=l+1}^NM_{r,T}})\text{N}(\eta_{l,T};\phi\eta_{l,T-1},1-\phi^2)\left(1-\exp\left(-\frac{\zeta_l^2+\eta_{l,T}^2}{2\alpha}\right)\right)^{M_{l,T}},\]
which is proportional to
\begin{multline*} \text{N}\left(\eta_{l,T};\frac{\phi\alpha\eta_{l,T-1}}{\alpha+\sum_{r=l+1}^NM_{r,T}-\phi^2\sum_{r=l+1}^NM_{r,T}},\frac{\alpha(1-\phi^2)}{\alpha+\sum_{r=l+1}^NM_{r,T}-\phi^2\sum_{r=l+1}^NM_{r,T}}\right)\\
\left(1-\exp\left(-\frac{\zeta_l^2+\eta_{l,T}^2}{2\alpha}\right)\right)^{M_{l,T}}
\end{multline*}
The slice samplers for $\eta_{l,1}$ and $\eta_{l,T}$ are therefore implemented in the same way as for $\eta_{l,t}$, except the normals which are sampled from have different means and variances.

\subsubsection*{Updating $\alpha$}

The full conditional for $\alpha$ is 
{\small \[p(\alpha\mid \dots,\text{data})\propto p(\alpha)\exp\left(-\frac{\sum_{t=1}^TM_{N,t}\sum_{l=1}^{N-1}(\zeta_l^2+\eta_{l,t}^2)}{2\alpha}\right)\exp\left(-\frac{\sum_{t=1}^T\sum_{l=2}^{N-1}M_{l,t}\sum_{r=1}^{l-1}(\zeta_r^2+\eta_{r,t}^2)}{2\alpha}\right)\]\[\prod_{t=1}^T\prod_{l=1}^{N-1}\left(1-\exp\left(-\frac{\zeta_l^2+\eta_{l,t}^2}{2\alpha}\right)\right)^{M_{l,t}}\]}

Therefore, with $p(\alpha)=\text{IG}(a_\alpha,b_\alpha)$, we have
\[p(\alpha\mid \dots,\text{data})=\text{IG}\left(\alpha;a_\alpha,b_\alpha+\frac{1}{2}\sum_{t=1}^T\left(M_{N,t}\sum_{l=1}^{N-1}(\zeta_l^2+\eta_{l,t}^2)+\sum_{l=2}^{N-1}M_{l,t}\sum_{r=1}^{l-1}(\zeta_r^2+\eta_{r,t}^2)\right)\right)\]
\[\prod_{t=1}^T\prod_{l=1}^{N-1}\left(1-\exp\left(-\frac{\zeta_l^2+\eta_{l,t}^2}{2\alpha}\right)\right)^{M_{l,t}}\]
The parameter $\alpha$ can be sampled using a Metropolis-Hastings algorithm. In particular we work with $\log(\alpha)$, and use a normal proposal distribution centered at the log of the current value of $\alpha$.

\subsubsection*{Updating $\phi$}
The full conditional for the AR parameter $\phi$ is 
\[p(\phi\mid \dots,\text{data})\propto p(\phi)\prod_{t=2}^T\prod_{l=1}^{N-1}\text{N}(\eta_{l,t};\phi\eta_{l,t-1},1-\phi^2)\]
\[\propto (1-\phi^2)^{-(N-1)(T-1)/2}\exp\left(-\sum_{t=2}^T\sum_{l=1}^{N-1}\frac{1}{2(1-\phi^2)}(\eta_{l,t}-\phi\eta_{l,t-1})^2\right)p(\phi)\]
We assume $p(\phi)=\mathrm{uniform}(0,1)$ or $p(\phi)=\mathrm{uniform}(-1,1)$, and apply a Metropolis-Hastings algorithm to sample $\log\left(\frac{\phi}{1-\phi}\right)$ or $\log\left(\frac{\phi+1}{1-\phi}\right)$, respectively, using a normal proposal distribution.

\subsubsection*{Updating $\{\boldsymbol{\mu}_{l,t}\}$}
The updates for $\boldsymbol{\mu}_{l,t}$ are $\text{N}(\boldsymbol{m}^*,\boldsymbol{V}^*)$, with $\boldsymbol{m}^*$ and $\boldsymbol{V}^*$ given by:
\begin{itemize}
\item For $t=2,\dots,T-1$, if $M_{l,t}=0$, then the update for $\boldsymbol{\mu}_{l,t}$ has $\boldsymbol{V}^*=(\boldsymbol{V}^{-1}+(\Theta^{-1}\boldsymbol{V}\Theta^{-T})^{-1})^{-1}$ and $\boldsymbol{m}^*=\boldsymbol{V}^*(\boldsymbol{V}^{-1}(\boldsymbol{m}+\Theta\boldsymbol{\mu}_{l,t-1})+(\Theta^{-1}\boldsymbol{V}\Theta^{-T})^{-1}\Theta^{-1}(\boldsymbol{\mu}_{l,t+1}-\boldsymbol{m}))$

\item For $t=2,\dots,T-1$, if $M_{l,t}\neq0$, then the update for $\boldsymbol{\mu}_{l,t}$ has $\boldsymbol{V}^*=(\boldsymbol{V}^{-1}+(\Theta^{-1}\boldsymbol{V}\Theta^{-T})^{-1}+M_{l,t}\boldsymbol{\Sigma}_l^{-1})^{-1}$ and $\boldsymbol{m}^*=\boldsymbol{V}^*(\boldsymbol{V}^{-1}(\boldsymbol{m}+\Theta\boldsymbol{\mu}_{l,t-1})+(\Theta^{-1}\boldsymbol{V}\Theta^{-T})^{-1}\Theta^{-1}(\boldsymbol{\mu}_{l,t+1}-\boldsymbol{m})+\boldsymbol{\Sigma}_l^{-1}\sum_{\{i:L_{t,i}=l\}}\boldsymbol{y}_{t,i})$

\item for $t=1$, if $M_{l,1}=0$, then the update for $\boldsymbol{\mu}_{l,1}$ has $\boldsymbol{V}^*=((\Theta^{-1}\boldsymbol{V}\Theta^{-T})^{-1}+\boldsymbol{V}_0^{-1})^{-1}$, and $\boldsymbol{m}^*=\boldsymbol{V}^*((\Theta^{-1}\boldsymbol{V}\Theta^{-T})^{-1}\Theta^{-1}(\boldsymbol{\mu}_{l,2}-\boldsymbol{m})+\boldsymbol{V}_0^{-1}\boldsymbol{m_0})$

\item for $t=1$, if $M_{l,1}\neq0$, then the update for $\boldsymbol{\mu}_{l,1}$ has $\boldsymbol{V}^*=(M_{l,1}\boldsymbol{\Sigma}_{l}^{-1}+(\Theta^{-1}\boldsymbol{V}\Theta^{-T})^{-1}+\boldsymbol{V}_0^{-1})^{-1}$ and $\boldsymbol{m}^*=\boldsymbol{V}^*(\boldsymbol{\Sigma}_l^{-1}\sum_{\{i:L_{1,i}=l\}}\boldsymbol{y}_{1,i}+(\Theta^{-1}\boldsymbol{V}\Theta^{-T})^{-1}\Theta^{-1}(\boldsymbol{\mu}_{l,2}-\boldsymbol{m})+\boldsymbol{V}_0^{-1}\boldsymbol{m_0})$

\item for $t=T$, if $M_{l,T}=0$, then the update for $\boldsymbol{\mu}_{l,T}$ has $\boldsymbol{V}^*=\boldsymbol{V}$, and $\boldsymbol{m}^*=\boldsymbol{m}+\Theta\boldsymbol{\mu}_{l,T-1}$
\item for $t=T$, if $M_{l,T}\neq 0$, then the update for $\boldsymbol{\mu}_{l,T}$ has $\boldsymbol{V}^*=(M_{l,T}\boldsymbol{\Sigma}_l^{-1}+\boldsymbol{V}^{-1})^{-1}$ and $\boldsymbol{m}^*=\boldsymbol{V}^*(\boldsymbol{\Sigma}_l^{-1}\sum_{\{i:L_{T,i}=l\}}\boldsymbol{y}_{t,i}+\boldsymbol{V}^{-1}(\boldsymbol{m}+\Theta\boldsymbol{\mu}_{l,T-1}))$
\end{itemize}

\subsubsection*{Updating $\{\boldsymbol{\Sigma}_{l}\}$}
The posterior full conditional for $\boldsymbol{\Sigma}_l$ is proportional to $\text{IW}(\nu+M_l,\boldsymbol{D}+\sum_{\{(t,i):L_{ti}=l\}}(\boldsymbol{y}_{t,i}-\boldsymbol{\mu}_l)(\boldsymbol{y}_{t,i}-\boldsymbol{\mu}_l)^T)$. When $|\{(t,i):L_{t,i}=l\}|=0$, $\boldsymbol{\Sigma}_l$ is drawn from $\text{IW}(\nu,\boldsymbol{D})$.

\end{document}